\documentclass{ptephy_v1}

\preprintnumber{XXXX-XXXX} 

\usepackage{amsmath} 
\usepackage{amsthm} 
\usepackage{graphics} 
\usepackage{bm}  
\usepackage{braket}
\usepackage{mathrsfs}
\usepackage{ulem}
\usepackage{indentfirst}

\begin{document}

\title{Non-Hermiticity and topological invariants of magnon Bogoliubov-de Gennes systems}


\author[1]{Hiroki Kondo}

\author[1]{Yutaka Akagi}

\author[1,2,3]{Hosho Katsura}

\affil[1]{Department of Physics, Graduate School of Science, The University of Tokyo, 7-3-1 Hongo, Tokyo 113-0033, Japan \email{kondo-hiroki290@g.ecc.u-tokyo.ac.jp}}
\affil[2]{Institute for Physics of Intelligence, The University of Tokyo, 7-3-1 Hongo, Tokyo 113-0033, Japan }
\affil[3]{Trans-scale Quantum Science Institute, The University of Tokyo, 7-3-1 Hongo, Tokyo 113-0033, Japan }


\begin{abstract}%
Since the theoretical prediction and experimental observation of the thermal Hall effect of magnons, a variety of novel phenomena that may occur in magnonic systems have been proposed. In this paper, we review the recent advances in the study of topological phases of magnon Bogoliubov-de Gennes (BdG) systems. After giving an overview of the previous works on electronic topological insulators and the thermal Hall effect of magnons, we provide the necessary background for bosonic BdG systems, with a particular emphasis on their non-Hermiticity arising from the diagonalization of the BdG Hamiltonian. After that, we introduce the definitions of $ \mathbb{Z}_2 $ topological invariants for bosonic systems with pseudo-time-reversal symmetry, which ensures the existence of bosonic counterparts of ``Kramers pairs". Because of the intrinsic non-Hermiticity of the bosonic BdG systems, these topological invariants have to be defined in terms of the bosonic Berry connection and curvature. We then introduce theoretical models that can be thought of as magnonic analogs of two- and three-dimensional topological insulators in class AII. We demonstrate analytically and numerically that the $ \mathbb{Z}_2 $ topological invariants precisely characterize the presence of gapless edge/surface states. We also predict that bilayer CrI$_3$ with a particular stacking would be an ideal candidate for the realization of a two-dimensional magnon system characterized by a nontrivial $ \mathbb{Z}_2 $ topological invariant. For three-dimensional topological magnon systems, the thermal Hall effect of magnons is expected to occur when a magnetic field is applied to the surface.
\end{abstract}

\subjectindex{xxxx, xxx}

\maketitle

\section{Introduction}

Our understanding of states of matter has developed mostly within the framework of the Ginzburg-Landau theory of symmetry breaking~\cite{Ginzburg50}.
In this framework, different phases are distinguished by local order parameters. 
On the other hand, there are states of matter beyond this successful paradigm such as integer/fractional quantum Hall systems~\cite{Klitzing80,Thouless82,Kohmoto85,Tsui82,Laughlin83,Hatsugai93a,Hatsugai93b,Haldane88}.
In particular, integer quantum Hall systems are classified by integer numbers related to beautiful mathematical concepts, i.e., topological invariants.
From a modern point of view, such systems fall into the category of topological insulators~\cite{Schnyder08,Kitaev09,Ryu10,Chiu13,Hasan10,Qi11} which have recently attracted considerable attention.
A characteristic feature of topological insulators and superconductors is the presence of gapless edge modes in their bulk energy gap that exhibit a variety of fascinating phenomena.
One of the important examples is the above quantization of Hall conductivity which is used as a standard of resistance.
Other notable examples include topological magnetoelectric effects in three-dimensional (3D) topological insulators~\cite{Qi08,Qi09,Nomura11,Morimoto16},
and realization of Majorana fermions~\cite{Mourik18,Kasahara18}, which is expected to be useful for robust quantum computation~\cite{Alicea12,DasSarma15,Aasen16}.

The research area of topological phases is not limited to electronic (or, more generally, fermionic) systems.
A number of intriguing phenomena such as the intrinsic thermal Hall effect have been studied  in bosonic systems such as magnons~\cite{Fujimoto09, Katsura10, Matsumoto11a,Matsumoto11b, Matsumoto14, Shindou13a, Shindou13b, Kim16, Onose10, Ideue12,Chisnell15, Hirschberger15, Han_Lee17,  Murakami_Okamoto17,Kawano19a,Kawano19b,Kawano19c,Cheng16,Seshadri18,Mook18,Mook14,Fransson16,Owerre17,Pershoguba18,Mook16,Li16,Su17,Nakata17a,Kim19,Owerre16a,Owerre16b,Wang17,Wang18}, photons~\cite{Onoda04, Hosten08, Raghu08, Haldane08, Wang09a, Ben-Abdallah16}, phonons~\cite{Strohm05, Sheng06a, Inyushkin07, Kagan08, Wang09b, Zhang10, Qin12, Mori14, Sugii17,Huber16}, and triplons~\cite{Rumhanyi15,Joshi17,Joshi19,Nawa19}.
Symmetry protected topological phases of bosons have also been proposed, for instance, in antiferromagnets~\cite{Zyuzin16}. Since such systems described by a bosonic Bogoliubov-de Gennes (BdG) Hamiltonian are intrinsically non-Hermitian, they do not fit into the topological classification of Hermitian systems~\cite{Matsumoto14}.
This implies that the topological invariants for bosons are not necessarily the same as those for electrons (more generally, fermions). We will indeed see that the standard definitions of Berry connection and curvature for fermions have to be modified when dealing with bosons.

This review focuses on the recent advances in the study of topological bosonic BdG systems, including some new results.
The organization of this paper is as follows.
In Sec.~\ref{previous}, we briefly review the previous studies on the topological properties of electron systems.
We also discuss the magnon thermal Hall effect, which is the initiation of the topological physics of magnon systems.
Section~\ref{Key} details the role of the pseudo-time-reversal operator which ensures the existence of Kramers pairs. In this section, we also show how the non-Hermiticity arises naturally in bosonic BdG systems and the resulting classification.
In Sec.~\ref{magnon_2D_3D}, we review recent progress in the ${\mathbb Z}_{2}$ topological phases of BdG systems in 2D and 3D.
We also propose candidate materials realizing the 2D magnonic ${\mathbb Z}_{2}$ topological phases. 
Section~\ref{summary} is devoted to a summary and future directions. 
In Appendix A and B, we provide some proofs and technical details of the results used in the main text.

\section{Previous studies on topological insulators of electrons and magnon thermal Hall effect\label{previous}}
In this section, we briefly review the previous studies on topological phases for both fermions and bosons.
In Sec.~\ref{Hall}, we first review the earlier studies on topological aspects of materials, the most notable example of which is the quantum Hall effect characterized by the first Chern number.
Sections~\ref{e_SHE} and \ref{e_3DTI} touch on the extension of the concept of quantum Hall systems, i.e., the ${\mathbb Z}_{2}$ topological insulators for fermions in 2D and 3D.
In Sec.~\ref{MTHE}, we introduce the magnon thermal Hall effect which is the bosonic counterpart of the quantum Hall effect for fermion.

\subsection{Quantum Hall effect and Chern number\label{Hall}}
\begin{figure}[!h]
\centering
  \includegraphics[width=0.4\columnwidth]{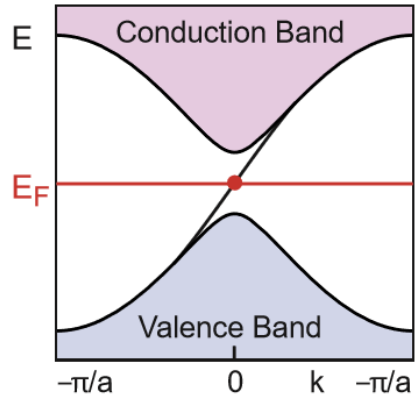}
\caption{The band structure of a semi-infinite strip of the Haldane model~\cite{Haldane88}, adopted from Ref.~\cite{Hasan10}. Shown in red is a chiral edge state across the gap between the valence and the conduction bands, which is responsible for the quantum Hall effect.
}\label{fig:chiraledge}
\end{figure}

Nontrivial topology of a quantum mechanical wave function results in, for example, the existence of surface states in a system with boundaries.
The initial study on topological phenomena in electronic systems can be traced back to the observation of the quantum Hall effect in which the Hall conductance is exactly quantized to an integer multiple of $e^2/h$~\cite{Klitzing80}.
The quantized Hall conductance is associated with the topology of the band structure by the 
Thouless-Kohmoto-Nightingale-den Nijs formula~\cite{Thouless82,Kohmoto85}.
By using the Kubo formula, the expression of the Hall conductance is obtained~\cite{Thouless82,Kohmoto85} as
\begin{align}
\sigma_{xy}=-\frac{e^{2}}{2\pi h}\sum_{E_n \le E_{\rm F}}\int_{\rm BZ} dk_{x}dk_{y}\Omega_{n}^{z}(\bm{k}),\label{eq:TKNN}
\end{align}
where $\Omega_{n}^{z}(\bm{k})= 2{\rm Im}\braket{\partial_{k_{x}}\psi_{n}(\bm{k})|\partial_{k_{y}} \psi_{n}(\bm{k})}$ ($\bm{k}=(k_{x},k_{y})$) is the Berry curvature of the $n$th band with energy $E_n (\bm{k})$  whose wave function is given by $\ket{\psi_{n}(\bm{k})}$ and BZ means Brillouin zone. 
The summation is taken over the bands below the Fermi level $E_{\rm F}$.
The topology of the band structure of a quantum Hall insulator is characterized by an integer, i.e., the Chern number:
\begin{align}
{\rm Ch}=\frac{1}{2\pi}\sum_{E_n \le E_{\rm F}}\int_{\rm BZ} dk_{x}dk_{y}\Omega_{n}^{z}(\bm{k}).
\end{align}
This corresponds to the number of the chiral edge states (the bulk-edge correspondence).
A schematic picture of the chiral edge state of a quantum Hall insulator is shown in Fig.~\ref{fig:chiraledge}.
As is seen from the band structure in Fig.~\ref{fig:chiraledge}, in order for the chiral edge states to exist, breaking time-reversal symmetry is necessary.
\noindent

\subsection{2D topological insulators of electrons \label{e_SHE}}

The bulk-edge correspondence is not limited to quantum Hall insulators, i.e. systems without time-reversal symmetry.
Time-reversal symmetry and other discrete symmetries lead to a variety of new topological phases, which are protected as long as such symmetries are preserved~\cite{Schnyder08,Kitaev09,Ryu10,Chiu13, Morimoto13, Fang12, Alexandradinata14, Shiozaki15a, Liu14, Fang15, Shiozaki15b, Shiozaki16, Wang16, Shiozaki17, Kruthoff17, Po17, Bradlyn17, Watanabe17}.

\begin{figure}[!h]
\centering
  \includegraphics[width=0.6\columnwidth]{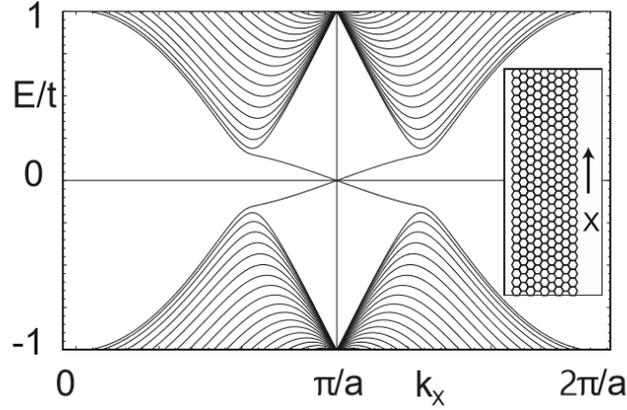}
\caption{The band structure of a strip of the Kane-Mele model with zigzag edges. 
A gapless helical edge state appears at $k_{x}=\pi/a$.
This figure is taken from Ref.~\cite{Kane05a}.
}\label{fig:helicaledge}
\end{figure}

The seminal examples of topological phases protected by time-reversal symmetry are $\mathbb{Z}_2$ topological insulators in class AII~\cite{Kane05a,Kane05b, Hasan10, Qi11}. 
The topological insulators in 2D possess a helical edge state which carries electrons with opposite spins propagating in opposite directions, resulting in the quantum spin Hall effect. 
A numerical result of the band structure calculation of a quantum spin Hall insulator is shown in Fig.~\ref{fig:helicaledge}.
The presence of the helical edge state is characterized by the $\mathbb{Z}_2$ index.
A model of a topological insulator with a nontrivial $\mathbb{Z}_2$ index was theoretically proposed by Kane and Mele by combining two copies of the Haldane model~\cite{Haldane88} so that the total system restores the time-reversal symmetry~\cite{Kane05a, Kane05b}.

The spin Hall insulator was first realized in HgTe/CdTe quantum well structures~\cite{Bernevig06,Konig07}, in which HgTe is sandwiched between the layers of CdTe.
When the thickness of the quantum well is $d<d_{c}=6.3$ nm, the system is a trivial insulator.
However, triggered by the band inversion for $d > d_c$, the $\mathbb{Z}_2$ topological insulator is realized in the layered material, exhibiting a quantized conductance.

Such topologically protected edge states can be understood as a Kramers pair, thus the Kramers theorem plays a crucial role in 2D topological insulators.
This theorem ensures that two degenerate states, i.e., Kramers pair, exist at the time-reversal-invariant momenta (TRIM) in electronic systems with  time-reversal symmetry. 

There are various expressions of $\mathbb{Z}_2$ indices for 2D topological insulators~\cite{Fu06,Fu07,Fukui07}.
One of them is given by integrating the Berry connection and curvature in the effective Brillouin zone (EBZ)~\cite{Fu06}.
The EBZ related to the time-reversal-invariant band structures refers to a half of the Brillouin zone.
The topological invariant of $n$th band is defined as 
\begin{align}
D_{n}  := \frac{1}{2\pi} \left[ \oint_{\partial {\rm EBZ}}  d\bm{k} \cdot \bm{A}_{n}(\bm{k})-\int_{\rm EBZ}dk_{x}dk_{y}\Omega_{n}^{z}(\bm{k}) \right]\hspace{1mm}{\rm mod}\hspace{1mm}2, \label{eq:e_z2}.
\end{align}
Here $\bm{A}_{n}(\bm{k})$ and $\Omega_{n}^{z}(\bm{k})$ are the summations over the Berry connection and curvature of degenerate states in the $n$th band:
\begin{align}
\bm{A}_{n}(\bm{k})=\sum_{l=1,2}\bm{A}_{n,l}(\bm{k}), \\
\Omega_{n}^{z}(\bm{k})=\sum_{l=1,2}\Omega_{n,l}^{z}(\bm{k}),
\end{align}
where
\begin{align}
&\bm{A}_{n,l}(\bm{k})={\rm i} \braket{\bm{\psi}_{n,l}(\bm{k})|\nabla_{\bm{k}}\bm{\psi}_{n,l}(\bm{k})}, \\
&\Omega_{n,l}^{z}(\bm{k})=\left(\nabla_{\bm{k}} \times \bm{A}_{n,l}(\bm{k})\right)_{z}.
\end{align}
Here, $\ket{\bm{\psi}_{n,l}(\bm{k})}$ is an eigenvector of a target Hamiltonian $H(\bm{k})$.
The index $l=1$ or $2$ denotes two degenerate states related by time-reversal operator $\Theta$, i.e., two states $\ket{\bm{\psi}_{n,1}(\bm{k})}$ and $\ket{\bm{\psi}_{n,2}(\bm{k})}$ satisfy $\ket{\bm{\psi}_{n,2}(\bm{k})}=-\Theta\ket{\bm{\psi}_{n,1}(-\bm{k})}$.
The summation over the bands under the Fermi level $E_{\rm F}$, namely $D = \sum_{E_n \le E_{\rm F}}  D_n$ (mod $2$),  corresponds to the number of gapless edge states across the energy gap in which $E_{\rm F}$ lies.
Such correspondence is discussed in Ref.~\cite{Fu06} by relating 2D topological insulator with a 1D spin pump.

\subsection{3D topological insulator of electrons\label{e_3DTI}}
After the proposal of the Kane-Mele model, the topological characterization of spin Hall insulators has been extended to 3D systems~\cite{Fu07,Moore07,Roy09,Guo09,Weeks10}.
The topological invariants for 3D topological insulators are defined as the winding numbers in the six EBZ in the 3D Brillouin zone, and written as follows:
\begin{align}
&\nu_{i,0}^{n} := \frac{1}{2\pi} \! \left[ \oint_{\partial {\rm EBZ}_{i,0}}  d\bm{k} \cdot \left[\bm{A}_{n}(\bm{k})\right]_{k_{i}=0}-\int_{{\rm EBZ}_{i,0}} dk_{j}dk_{k} \left[\Omega_{n}^{i}(\bm{k})\right]_{k_{i}=0} \right]\hspace{1mm}{\rm mod}\hspace{1mm}2, \label{eq:e_Z2_0} \\
&\nu_{i,\pi}^{n} := \frac{1}{2\pi}  \left[ \oint_{\partial {\rm EBZ}_{i,\pi}} d\bm{k} \cdot \left[\bm{A}_{n}(\bm{k})\right]_{k_{i}=\pi} -\int_{{\rm EBZ}_{i,\pi}} dk_{j}dk_{k}  \left[\Omega_{n}^{i}(\bm{k})\right]_{k_{i}=\pi} \right] \hspace{1mm}{\rm mod}\hspace{1mm}2, \label{eq:e_Z2_pi}
\end{align}
where $n$ is a band index and $i=x$, $y$, and $z$.
Here, $j$ and $k$ represent two of $x,y$, and $z$ which are different from $i$.
The notation ${\rm EBZ}_{x,0}$ $({\rm EBZ}_{x,\pi})$ stands for the effective Brillouin zone in the $k_{x}=0$ $(k_{x}=\pi)$ plane, which is specified as $k_{x}=0$ $(k_{x}=\pi),k_{y}\in[-\pi,\pi],k_{z}\in[0,\pi]$. Its boundary is denoted as $\partial {\rm EBZ}_{x,0}$ $({\rm EBZ}_{x,\pi})$. The other four effective Brillouin zones are defined similarly.
This is a 3D extension of the formula Eq.~(\ref{eq:e_z2}).
We write the winding numbers over the bands under the Fermi level as
\begin{align}
\nu_{i,0}:=\sum_{E_n \le E_{\rm F}}\nu_{i,0}^{n}\hspace{5mm}{\rm mod} \hspace{1mm}2,  \\
\nu_{i,\pi}:=\sum_{E_n \le E_{\rm F}}\nu_{i,\pi}^{n}\hspace{5mm}{\rm mod} \hspace{1mm}2 .
\end{align}
The winding number in the bulk system $\nu_{i,0}$ $(\nu_{i,\pi})$ counts the parity of the total number of surface Dirac cones on the $k_{i}=0$ $(k_{i}=\pi)$ line in the 2D Brillouin zone of the slab.
For example, we consider how the winding number $\nu_{x,0}$ is related to the surface states in a slab with a (001) face.
The slab breaks the translation symmetry in the $z$-direction. Thus, Fourier transformation cannot be applied in the $z$-direction.
The energy spectrum at a point on 2D Brillouin zone $-\pi\leq k_{x},k_{y}\leq\pi$ has a contribution from all $k_{z}$ in the bulk band structure.
Let us project a 3D Brillouin zone into the $k_{x}$-$k_{y}$ plane.
The effective Brillouin zone ${\rm EBZ}_{x,0}$ is mapped into a line $k_{x}=0,k_{y}\in [0,\pi]$ in the 2D Brillouin zone.
Since the topological invariant $\nu_{x,0}$ is defined as the winding number in ${\rm EBZ}_{x,0}$, we can see that $\nu_{x,0}$ counts the number of Dirac cones at the wave vectors $(k_x, k_y) = (0, 0), (0, \pi)$, which are obtained by projecting the four TRIM $\bm{k} = (0, 0, 0)$, $(0, \pi, 0)$, $(0, 0, \pi)$, $(0, \pi, \pi)$ in ${\rm EBZ}_{x,0}$ onto the 2D Brillouin zone.
Along the same lines, one can see that a similar correspondence holds for (001),(010), and (100) faces.

Since $\nu_{i,0}+\nu_{i,\pi}$ counts the total number of Dirac cones modulo $2$, one has
\begin{align}
\nu_{x,0}+\nu_{x,\pi}=\nu_{y,0}+\nu_{y,\pi}=\nu_{z,0}+\nu_{z,\pi} \hspace{5mm}{\rm mod} \hspace{1mm}2.
\end{align}
It follows from this equation that only four of the six topological invariants are independent.
Thus, the topological phase of the system is completely characterized by the set of the four topological invariants: $(\nu_{0};\nu_{x},\nu_{y},\nu_{z})$, where  
\begin{align}
&\nu_{0}=\nu_{x,0}+\nu_{x,\pi} \hspace{5mm}{\rm mod} \hspace{1mm}2, \\
&\nu_{i}=\nu_{i,\pi}\hspace{5mm}(i=x,y,z).
\end{align}
We note that one of the topological invariants, $\nu_{0}$ counts the parity of the total number of Dirac cones.
For $\nu_{0}=1$, there exists an odd number of Dirac cones in total. Such a topological phase, so-called the strong topological phase, is robust against disorder which does not break the time-reversal symmetry. 
If the four topological invariants are all zero, the system is in the trivial phase.
When $\nu_{0}=0$ and at least one of $\nu_{i}$ $(i=x,y,z)$ is nonzero, there exist an even number of Dirac cones in total.
However, this phase is not robust against disorder which does not break the time-reversal symmetry because an even number of Dirac cones annihilate each other by perturbation including disorder, which results in the opening of the band gap.
Thus, the phase is called the weak topological phase.

\begin{figure}[!h]
\centering
  \includegraphics[width=0.6\columnwidth]{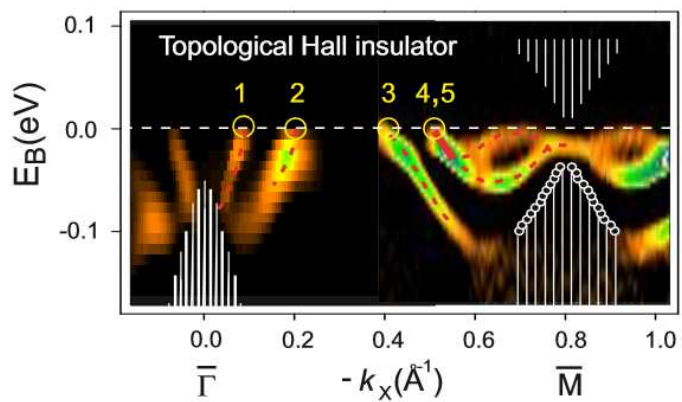}
\caption{ARPES measurement of the surface band structure of ${\rm Bi}_{0.9}{\rm Sb}_{0.1}$. 
The surface band crosses the Fermi surface five times between two time-reversal-invariant points ${\rm \bar{\Gamma}}$ and ${\rm \bar{M}}$ in the surface Brillouin zone.
This figure is taken from Ref.~\cite{Hsieh08}.
}\label{fig:3DTIARPES}
\end{figure}

The first 3D topological insulator identified experimentally is ${\rm Bi}_{1-x}{\rm Sb}_{x}$~\cite{Hsieh08}.
The unusual surface state of ${\rm Bi}_{1-x}{\rm Sb}_{x}$ is measured by an angle-resolved photoemission spectroscopy (ARPES) experiment (See Fig.~\ref{fig:3DTIARPES}).
\noindent
The strong topological phase is realized for $0.07 \leq x \leq 0.22$ while the band structure of ${\rm Bi}_{1-x}{\rm Sb}_{x}$ is complicated and the band gap is small.
On the other hand, the second generation topological insulators ${\rm Bi}_{2}{\rm Se}_{3},{\rm Bi}_{2}{\rm Te}_{3}$, and ${\rm Sb}_{2}{\rm Se}_{3}$ realize a single Dirac cone and a larger band gap~\cite{Zhang09,Xia09}.

\subsection{Magnons and thermal Hall effect\label{MTHE}}

So far we have discussed electronic (fermionic) topological insulators in 2D and 3D. However, the notion of topology is not limited to fermionic systems. In fact, the past two decades have also witnessed the role of topology in a variety of bosonic systems such as  magnons~\cite{Fujimoto09, Katsura10, Matsumoto11a,Matsumoto11b, Matsumoto14, Shindou13a, Shindou13b, Kim16, Onose10, Ideue12,Chisnell15, Hirschberger15, Han_Lee17,  Murakami_Okamoto17,Kawano19a,Kawano19b,Kawano19c,Cheng16,Seshadri18,Mook18,Mook14,Fransson16,Owerre17,
Pershoguba18,Mook16,Li16,Su17,Nakata17a,Kim19,Owerre16a,Owerre16b,Wang17,Wang18}, photons~\cite{Onoda04, Hosten08, Raghu08, Haldane08, Wang09a, Ben-Abdallah16}, phonons~\cite{Strohm05, Sheng06a, Inyushkin07, Kagan08, Wang09b, Zhang10, Qin12, Mori14, Sugii17,Huber16}, and triplons~\cite{Rumhanyi15,Joshi17,Joshi19,Nawa19}, which exhibit fascinating phenomena akin to the Hall effect.
Of particular interest in this review are magnons that are the quasiparticles of low-energy collective excitations in magnets.
Magnons could be observed in real-time/space in experiments and have potential applications in spintronics, as they have long coherence and carry angular momenta~\cite{Demokritov01}.
The topological phenomena in magnonic systems were initiated by the theoretical prediction of the thermal Hall effect of magnons by one of the authors and his collaborators~\cite{Katsura10}.

Historically, the concept of magnons was first introduced by Bloch in the 1930s~\cite{Bloch1930} to explain the reduction of spontaneous magnetization in ferromagnets.
For our purpose, it is convenient to introduce the mapping between spin operators and bosonic creation and annihilation operators called the Holstein-Primakoff transformation~\cite{Holstein40}.
To define this transformation, let us introduce some notation. Let $S^z_i$ be the $z$-component of the spin operator and $S^{\pm}_i = S^x_i \pm i S^y_i$ the spin raising/lowering operator at lattice site $i$. These operators can be written in terms of bosonic operators as
\begin{align}
&S_{i}^{z}=S-b_{i}^{\dagger}b_{i}, \\
&S_{i}^{+}=\sqrt{2S-b_{i}^{\dagger}b_{i}}b_{i}, \\
&S_{i}^{-}=b_{i}^{\dagger}\sqrt{2S-b_{i}^{\dagger}b_{i}}, \label{eq:HP}
\end{align}
where the operator $b_{i}$ annihilates a magnon at site $i$.
For the spin operator $\bm{S}_{i}$ to satisfy the commutation relations of angular momentum, the operator $b_{i}$ must satisfy the bosonic commutation relations $[b_{i},b^{\dagger}_{j}]=\delta_{ij}$.
Within the approximation of neglecting the interactions between magnons, the above formula simplifies to the following:
\begin{align}
&S_{i}^{z}=S-b_{i}^{\dagger}b_{i}, \\
&S_{i}^{+}\simeq \sqrt{2S}b_{i}, \\
&S_{i}^{-}\simeq \sqrt{2S}b_{i}^{\dagger}.
\end{align}
This approximation is valid when the spin magnitude $S$ is large and/or the temperature is low enough that the population of thermally activated magnons at each site is small.

Let us now see how magnetic interactions are expressed in terms of bosonic operators.
The Dzyaloshinskii-Moriya (DM) interaction $\bm{D}_{ij} \cdot (\bm{S}_{i} \times \bm{S}_{j})$ is an antisymmetric magnetic exchange interaction between two spins, which originates from the spin-orbit interaction~\cite{Dzyaloshinskii1960,Moriya1960a,Moriya1960b}. 
The Heisenberg and DM interactions correspond to the real and purely imaginary hopping terms of the magnon Hamiltonian, respectively. The complex phase factors arising from the combination of these two give rise to the nontrivial topology of the magnon wave functions, leading to the magnon thermal Hall effect.

\begin{figure}[!h]
\centering
  \includegraphics[width=0.5\columnwidth]{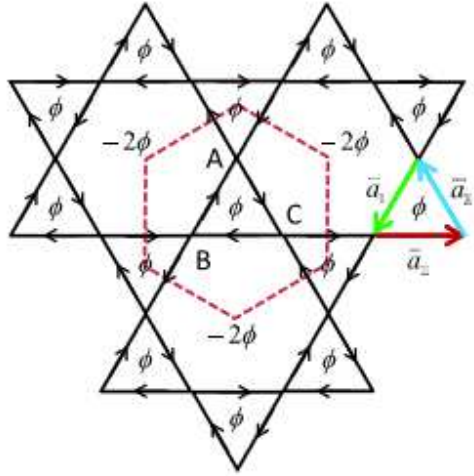}
\caption{Kagome lattice system with fictitious fluxes experienced by magnons which come from the DM interaction (or the scalar spin chirality term). Magnons acquire the phase factor $e^{i\phi}$ $(e^{-2i\phi})$ by going around a triangle (hexagon) in the clockwise (counter-clockwise) direction. This gives rise to a nonzero Berry curvature of magnons, leading to the thermal Hall effect. 
This figure is taken from Ref.~\cite{Katsura10}.
}\label{fig:Katsura10}
\end{figure}

\begin{figure}[!h]
\centering
  \includegraphics[width=0.6\columnwidth]{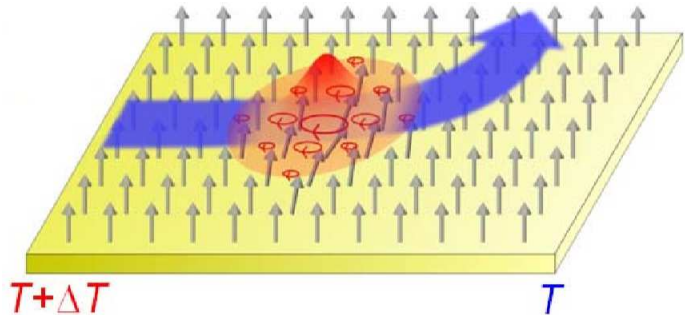}
\caption{Schematic picture of the magnon thermal Hall effect. 
The magnetic field is applied in the $z$-direction.
A magnon wave packet in the ferromagnet moving from the hot to the cold side ($x$-direction) drifts in the $y$-direction by the DM interaction.
This figure is taken from Ref.~\cite{Onose10}.
}\label{fig:MTHE}
\end{figure}

The magnon thermal Hall effect has been predicted theoretically in the kagome lattice ferromagnet with a scalar spin chirality term~\cite{Katsura10} which plays essentially the same role as the DM interaction. 
Within the above approximation, the scalar chirality term and the DM interaction result in the same purely imaginary hopping term in the magnon Hamiltonian.
Figure~\ref{fig:Katsura10} shows the pattern of fictitious fluxes experienced by magnons in a kagome ferromagnet, which result from the DM interaction (or scalar spin chirality term).
\noindent
The schematic picture of the magnon thermal Hall effect is shown in Fig.~\ref{fig:MTHE}.
While magnons are chargeless particles that are unaffected by electric fields, magnon current can be induced by applying a temperature gradient.
Magnon Hall current conveys energy in the direction perpendicular to both temperature gradient and the magnetic field.
\noindent
Using semiclassical analysis and linear response theory, Matsumoto and Murakami pointed out that the expression of the thermal Hall coefficient derived in Ref.~\cite{Katsura10} lacks the term of orbital angular momentum of magnons~\cite{Matsumoto11a}.
The modified expression of the thermal Hall coefficient is written as follows:
\begin{align}
\kappa^{xy}=\frac{k_{B}^{2}T}{\hbar V}\sum_{n,\bm{k}}c_{2}(\rho_{n})\Omega_{n}^{z}(\bm{k}),\label{eq:thermal_Hall_coefficient}
\end{align}
where $\Omega_{n}^{z}(\bm{k})=2{\rm Im} \left<\partial_{k_{x}}\psi_{n}(\bm{k})|\partial_{k_{y}}\psi_{n}(\bm{k})\right>$ is the Berry curvature of the $n$th magnon band.
Here $\ket{\psi_{n}(\bm{k})}$ is the $n$th eigenvector of the magnon Hamiltonian in $\bm{k}$-space.

The function $c_{2}(\rho)$ is defined as $c_{2}(\rho)=(1+\rho)\left(\log\frac{1+\rho}{\rho}\right)^{2}-\left(\log\rho\right)^{2}-2{\rm Li}_{2}(-\rho)$, where ${\rm Li}_{n}(x)$ is the polylogarithm function.
Clearly, in the same way as electrons, the thermal Hall coefficient given by Eq.~(\ref{eq:thermal_Hall_coefficient}) is described by the Berry curvature.
The correspondence between the Chern number
\begin{align}
{\rm Ch}_{n}=\frac{1}{2\pi}\int_{\rm BZ}d\bm{k}\Omega_{n}(\bm{k})
\end{align}
and the number of gapless edge states of magnons is confirmed~\cite{Matsumoto11a}.
Some comments are in order here. Although the definition of the Chern number for magnons is exactly the same as the fermionic one, the thermal Hall coefficient is not quantized. This is because magnons obey Bose-Einstein statistics and filling their energy bands up to the ``Fermi level" does not make sense.
Another comment is that the above formula for the bosonic Chern number is valid only when the number of bosons is conserved.
In general, magnon (boson) systems described by the BdG-type Hamiltonian do not conserve the number of particles, and thus the expression of the Chern number is modified.
In addition, bosonic BdG systems have a unique non-Hermitian property.
This is because a general bosonic BdG Hamiltonian has to be diagonalized by a para-unitary matrix that preserves the cannonical bosonic commutation relations, which amounts to the diagonalization of a non-Hermitian matrix. 
As a consequence, the expressions of the Berry connection and curvature differ from those of electrons.
The details will be discussed in Sec.~\ref{boson_BdG}.

\begin{figure}[!h]
\centering
  \includegraphics[width=0.6\columnwidth]{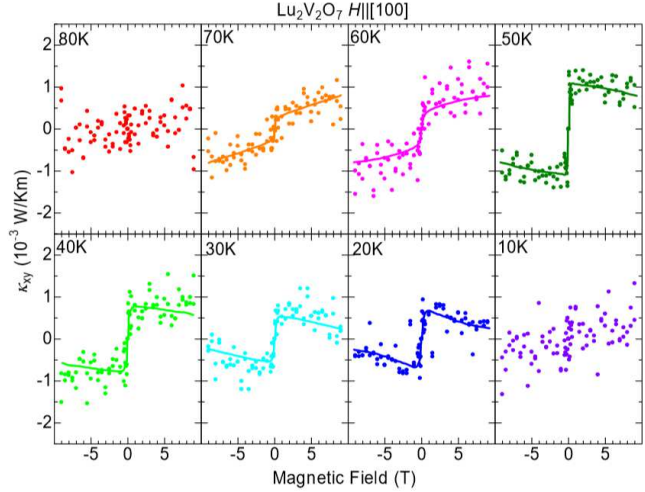}
\caption{Magnetic field variation of the thermal Hall conductivity of Lu$_2$V$_2$O$_7$ at various temperatures. This figure is taken from Ref.~\cite{Onose10}.
The magnetic transition temperature is known to be 70 K.
The magnon Hall effect is most clearly observed around 50 K. 
At 10 K and 80 K, the magnetic field dependence of the thermal Hall coefficient is unclear.
This would be because the number of magnons is small at too low temperature and thermal fluctuation is large at high temperatures.
}\label{fig:thermalHallconductivity}
\end{figure}

Although the original theoretical work was concerned with 2D systems, the magnon thermal Hall effect was first observed in a 3D pyrochlore ferromagnet Lu$_2$V$_2$O$_7$~\cite{Onose10}.
The underlying mechanism of the effect is, however, essentially the same as the one for 2D. 
Figure~\ref{fig:thermalHallconductivity} shows the experimental results of the thermal Hall conductivity of the ferromagnetic Mott insulator Lu$_2$V$_2$O$_7$.
In this material, the $S=1/2$ magnetic moments carried by ${\rm V}^{4+}$ ions form a network of corner sharing tetrahedra, i.e., a pyrochlore lattice. 
The system has only magnons (and phonons) as mobile quasi-particles because it is a Mott insulator. Therefore, the result indicates that magnons contribute to the observed thermal Hall effect. The solid curves in Fig. 6 are the theoretical fitting curves. Clearly, the theory accounts well for the experimental data (See Ref.~\cite{Onose10} for a more detailed discussion).

Recent theoretical works predict that the magnon thermal Hall effect occurs in a Kitaev material at a high magnetic field~\cite{McClarty18,Joshi18,Cookmeyer18,Lu18}.
The nonzero Berry curvature of magnons in this model is induced not by the DM interactions but by the off-diagonal symmetric exchange interactions called the $ \Gamma $ terms.

\noindent

\section{Non-Hermiticity and Symmetries of bosonic BdG Hamiltonians\label{Key}}


In this section, we review the mathematical background of bosonic BdG systems.
In Sec.~\ref{boson_BdG}, we show how to diagonalize a bosonic BdG Hamiltonian and define the Berry connection and curvature in terms of the eigenvectors of a non-Hermitian matrix arising from the BdG Hamiltonian.
Due to the non-Hermiticity of bosonic BdG systems, their definitions are different from those of electrons.
Section~\ref{PTRS} provides the pseudo-time-reversal operator which plays an important role in ensuring the existence of bosonic ``Kramers pairs"~\cite{Wu15,Ochiai15,He16}.
In Sec.~\ref{Kramers_boson}, we prove the existence of ``Kramers pairs" in a system with the pseudo-time-reversal symmetry.
In Sec.~\ref{Ham_PTRS}, we show how the pseudo-time-reversal symmetry restricts the form of the Hamiltonian.
In Sec.~\ref{NHPT}, we review the topological classification of non-Hermitian systems including bosonic BdG ones. We also touch on topological bosonic phases and their classes.

\subsection{Diagonalization of bosonic BdG Hamiltonian\label{boson_BdG}}

We shall show how to obtain the band structure and eigenstates by diagonalizing the BdG Hamiltonian~\cite{Colpa78,Gurarie03} by a para-unitary matrix. We follow the approach of Ref.~\cite{Colpa78}.
At the end of the section, we give the expressions for the Berry connection and curvature of the system.
We begin with the bosonic BdG Hamiltonian in $k$-space
\begin{align}
&\mathcal{H}=\frac{1}{2}\sum_{\bm{k}}\bm{\phi}^{\dagger}(\bm{k}) H(\bm{k})\bm{\phi}(\bm{k}), \label{eq:Ham}\\
&\bm{\phi}^\dagger ({\bm k}) = [ \beta^\dagger_1({\bm k}), \cdots, \beta^\dagger_{\mathscr{N}} ({\bm k}), \beta_1 (-{\bm k}), \cdots, \beta_{\mathscr{N}} (-{\bm k}) ].\label{eq:op_phi}
\end{align}
Here, $\bm{\beta}^{\dagger}(\bm{k})=[\beta_{1}^{\dagger}(\bm{k}),\cdots,\beta_{\mathscr{N}}^{\dagger}(\bm{k})]$ denotes boson creation operators with momentum $\bm{k}$.
The subscript $\mathscr{N}$ is the number of internal degrees of freedom in a unit cell.
The matrix $H(\bm{k})$ is written as
\begin{align}
H(\bm{k})=\left(
\begin{array}{cc}
h(\bm{k}) & \Delta(\bm{k}) \\
\Delta^{*}(-\bm{k})&h^{*}(-\bm{k})  \\
\end{array} 
\right).
\end{align}
Since $H(\bm{k})$ is Hermitian,  $h(\bm{k})$ and $\Delta(\bm{k})$ satisfy $h(\bm{k})=h^{\dagger}(\bm{k})$ and $\Delta^{T}(\bm{k})=\Delta(-\bm{k})$, respectively.
The components of the operator $\bm{\phi}(\bm{k})$ satisfy the commutation relation $[\phi_{i}(\bm{k}),\phi_{j}^{\dagger}(\bm{k}')]=(\Sigma_{z})_{ij}\delta_{\bm{k},\bm{k}'}$.
Here, $\Sigma_{z}$ is defined as a tensor product $\Sigma_{z}:=\sigma_z \otimes  1_\mathscr{N}$, where $\sigma_a$ ($a=x,y,z$) is the $a$-component of the Pauli matrix acting on the particle-hole space and $1_\mathscr{N}$ is the $\mathscr{N} \times \mathscr{N}$ identity matrix. 

Let us look for conditions under which the transformation matrix $T(\bm{k})$ leaves the bosonic commutation relation unchanged  (Such a matrix is called a para-unitary matrix.).
The commutator of $\psi_{i}(\bm{k})=(T^{-1}(\bm{k})\bm{\phi}(\bm{k}))_{i}$ and $\psi_{j}^{\dagger}(\bm{k})$ is written as
\begin{align}
[\psi_{i}(\bm{k}),\psi_{j}^{\dagger}(\bm{k})]
&=[(T^{-1}(\bm{k}))_{ik}\phi_{k}(\bm{k}),(T^{-1}(\bm{k}))_{jl}^{*}\phi_{l}^{\dagger}(\bm{k})]
=(T^{-1}(\bm{k}))_{ik}(\Sigma_{z})_{kl}(T^{-1}(\bm{k}))_{jl}^{*} \nonumber\\
&=\left(T^{-1}(\bm{k})\Sigma_{z}(T^{-1}(\bm{k}))^{\dagger}\right)_{ij},
\end{align}
where repeated indices are summed over. By requiring $[\psi_{i}(\bm{k}),\psi_{j}^{\dagger}(\bm{k}')]=(\Sigma_{z})_{ij}\delta_{\bm{k},\bm{k}'}$, we obtain the para-unitary condition:
\begin{align}
T(\bm{k})\Sigma_{z}T^{\dagger}(\bm{k})=\Sigma_{z}.
\label{eq:paraunitary}
\end{align}
Thus, we must diagonalize the BdG Hamiltonian by using a matrix satisfying the above para-unitarity~(\ref{eq:paraunitary}).

To identify the appropriate $T({\bm k})$, it is useful to note certain properties of $ \Sigma_z H ({\bm k}) $. Suppose that the matrix $H({\bm k})$ is positive definite, i.e., all eigenvalues are positive. Then, the following three statements hold: 
\begin{description}
  \item[(i)]The eigenvalues of the matrix $ \Sigma_z H({\bm k}) $ are real and nonzero.
  \item[(ii)]If $\bm{v}({\bm k}) $ is an eigenvector of $ \Sigma_z H ({\bm k}) $ with eigenvalue $E({\bm k}) $, then $\Sigma_x \bm{v}^{*} (-{\bm k}) $ is an eigenvector of $\Sigma_z H ({\bm k}) $ with eigenvalue $ -E({\bm k}) $, where $\Sigma_{x}$ is defined as a tensor product $\Sigma_{x}:=\sigma_x \otimes  1_\mathscr{N}$.
  \item[(iii)] By using the indices $n=1,\cdots, \mathscr{N}$ and $\sigma=\pm$, $2\mathscr{N}$ eigenvectors can be taken to satisfy para-orthogonality $\bm{v}_{n\sigma}^{\dagger}(\bm{k})\Sigma_{z}\bm{v}_{m\rho}(\bm{k})=\sigma\delta_{nm}\delta_{\sigma\rho}$.
\end{description}
The details of the proofs of these are shown in Appendix A.
By using (i),(ii), and (iii), one finds that the matrix defined as
\begin{align}
T(\bm{k})=\left(\bm{v}_{1+}(\bm{k}),\cdots,\bm{v}_{\mathscr{N}+}(\bm{k}), \bm{v}_{1-}(\bm{k}),\cdots,\bm{v}_{\mathscr{N}-}(\bm{k}) \right)
\end{align}
satisfies para-unitarity Eq.~(\ref{eq:paraunitary}), where the eigenvectors $\bm{v}_{n+}(\bm{k})$ and $\bm{v}_{n-}(\bm{k})$ are related by $\bm{v}_{n+}(\bm{k})=\Sigma_{x}\bm{v}_{n+}^{*}(-\bm{k})$ $(n=1,\cdots,\mathscr{N})$.
The matrix $T(\bm{k})$ diagonalizes the Hamiltonian $H(\bm{k})$:
\begin{align}
T^{\dagger}(\bm{k})H(\bm{k})T(\bm{k})={\rm diag}\left( E_{1}(\bm{k}),\cdots, E_{\mathscr{N}}(\bm{k}),E_{1}(-\bm{k}),\cdots, E_{\mathscr{N}}(-\bm{k}) \right).
\end{align}
Therefore, solving the eigenvalue problem $\Sigma_z H(\bm{k})\bm{v}(\bm{k}) = E(\bm{k})\bm{v}(\bm{k})$, we obtain the eigenvalues and the para-unitary matrix automatically. However, the matrix $\Sigma_z H(\bm{k})$ is no longer Hermitian, thus the bosonic BdG systems have to be handled within the framework of non-Hermitian quantum mechanics. The non-Hermiticity modifies the inner-product for
bosonic wave functions as
\begin{align}
\langle\langle\bm{\phi},\bm{\psi}\rangle\rangle=\bm{\phi}^{\dagger}\Sigma_z \bm{\psi},
\end{align}
where $\bm{\phi}$ and $\bm{\psi}$ are $2N$-dimensional complex vectors and $\bm{\phi}^{\dagger}$ is the adjoint of $\bm{\phi}$~\cite{Lein19}. Reflecting the non-trivial inner-product, the Berry connection and curvature of the bosonic systems described by BdG Hamiltonian are written as
\begin{align}
&\bm{A}_{n\sigma}(\bm{k})={\rm i}\sigma\bm{v}_{n\sigma}^{\dagger}(\bm{k})\Sigma_{z}\bm{v}_{n\sigma}(\bm{k}), \label{eq:BdG_boson_Berry_connection}\\
&\bm{\Omega}_{n\sigma}(\bm{k})=\nabla_{\bm{k}}\times \bm{A}_{n\sigma}(\bm{k}) \label{eq:BdG_boson_Berry}.
\end{align}
We refer the reader to Ref.~\cite{Matsumoto14} for the detailed derivation of these formulas.

\subsection{Pseudo-time-reversal symmetry\label{PTRS}}
As we mentioned in Sec.~\ref{previous}, the Kramers theorem plays an important role in the construction of quantum spin Hall insulators.
However, the Kramers theorem  cannot be directly applied to bosonic systems such as magnons.
In this section, in order to introduce the concept of Kramers' pair in bosonic systems, we define fermion-like symmetry dubbed pseudo-time-reversal symmetry, following Ref.~\cite{Kondo19a}.
Based on this symmetry,  $\mathbb{Z}_{2}$ topological invariants for magnonic (bosonic) systems will be defined in Sec.~\ref{magnon_2D_3D}.

The fermion-like pseudo-time-reversal operator in bosonic BdG systems is generally given by $\Theta'=PK$ where $P$ is a ${\bm k}$-independent para-unitary matrix and $K$ is the complex conjugation. 
The operator $\Theta'$ satisfies the following relation: 
\begin{align}
\Theta'^{2}=-1 \label{eq:Theta}.
\end{align}
By the operator $\Theta'$, we define pseudo-time-reversal symmetric systems which meet the following condition:
\begin{align}
\Sigma_{z}H(-\bm{k})\Theta'-\Theta'\Sigma_{z}H(\bm{k})=0, 
\label{eq:Commutation}
\end{align}
where the bosonic BdG Hamiltonian matrix $H(k)$ is given by Eq.~(\ref{eq:Ham}) and we assume the subscript $\mathscr{N}$ is even.
Note that the operator $\Theta'$ satisfies Eq.~(\ref{eq:Theta}) as in fermionic systems, while
the conventional time-reversal operator\footnote{Magnons are spin-1 bosonic particles. Thus the time-reversal operator $\Theta$ for magnonic systems must satisfy $\Theta^2=1$. See, for example, Ref.~\cite{Sakurai17}.} squares to $+1$ for bosonic systems. 
Explicit expressions for $\Theta'$ and $H({\bm k})$ will be given later in Eqs. (\ref{eq:pseudoTR}) and (\ref{eq:H}).

\subsection{Kramers pair of bosons\label{Kramers_boson}}

In this section, we show that 
the pseudo-time-reversal  operator $\Theta'$ ensures the existence of ``Kramers pairs" of bosons~\cite{Kondo19a}.
To begin with, let us consider the eigen-equation of the bosonic BdG Hamiltonian:
\begin{align}
\Sigma_{z}H(\bm{k})\bm{\psi}(\bm{k})=E(\bm{k})\bm{\psi}(\bm{k}). 
\label{eq:BdG_sup}
\end{align}
Multiplying both sides of Eq.~(\ref{eq:BdG_sup}) from the left by $\Theta'$, we obtain
\begin{align}
\Sigma_{z}H(-\bm{k})\Theta'\bm{\psi}(\bm{k})=E(\bm{k})\Theta'\bm{\psi}(\bm{k}),
\label{eq:BdG_TR}
\end{align}
where we used Eq.~(\ref{eq:Commutation}). 
From Eqs. (\ref{eq:BdG_sup}) and (\ref{eq:BdG_TR}) for  the time-reversal-invariant momenta (TRIM) $\bm{k}=\bm{\Lambda}$, we find that the two vectors $\bm{\psi}(\bm{\Lambda})$ and $\Theta'(\bm{\Lambda})\bm{\psi}(\bm{\Lambda})$ are eigenvectors of $\Sigma_{z}H(\bm{\Lambda})$ with the same eigenvalue $E(\bm{\Lambda})$.
In the following, we prove that these two vectors are 
orthogonal to each other. We first note that the inner product 
of $\bm{\phi}(-\bm{k})$ and $\Theta'\bm{\psi}(\bm{k})$
yields
\begin{align}
\left\langle \!  \left\langle \bm{\phi}(-\bm{k}),\Theta'\bm{\psi}(\bm{k})\right\rangle \! \right\rangle 
&=\phi_{i}^{*}(-\bm{k})\left(\Sigma_{z}P\right)_{ij}\psi_{j}^{*}(\bm{k}) \nonumber \\
&=\psi_{j}^{*}(\bm{k})\left(\Sigma_{z}P\right)_{ji}^{T}\phi_{i}^{*}(-\bm{k}) \nonumber \\
&=\left\langle \! \left\langle \bm{\psi}(\bm{k}), \Sigma_{z}P^{T}\Sigma_{z}K\bm{\phi}(-\bm{k})\right\rangle \! \right\rangle.  \label{eq:Kramers1}
\end{align}
By replacing $\bm{\phi}(-\bm{k})$ with $\Theta'\bm{\phi}(-\bm{k})$, 
the inner product can be cast into the following form:
\begin{align}
\left\langle \! \left\langle \Theta'\bm{\phi}(-\bm{k}),\Theta'\bm{\psi}(\bm{k})\right\rangle \! \right\rangle 
&=\left\langle \! \left\langle \bm{\psi}(\bm{k}), \Sigma_{z}P^{T}\Sigma_{z}KPK\bm{\phi}(-\bm{k})\right\rangle \! \right\rangle \nonumber \\
&=\left\langle \! \left\langle \bm{\psi}(\bm{k}), \bm{\phi}(-\bm{k})\right\rangle \! \right\rangle,
\label{eq:Kramers2}
\end{align}
where we used the para-unitary condition $P^{\dagger} \Sigma_z P = \Sigma_z$ and $(\Sigma_z)^2 = 1_{2 {\mathcal N}}$. Then one finds that the inner product of $\bm{\psi}(\bm{k})$ and $\Theta'\bm{\psi}(-\bm{k})$ satisfies
\begin{align}
\left\langle \! \left\langle \bm{\psi}(\bm{k}),\Theta'\bm{\psi}(-\bm{k})\right\rangle \!\right\rangle 
&=\left\langle \! \left\langle \Theta'^{2}\bm{\psi}(-\bm{k}), \Theta'\bm{\psi}(\bm{k})\right\rangle \! \right\rangle \nonumber \\
&=-\left\langle \! \left\langle \bm{\psi}(-\bm{k}),\Theta'\bm{\psi}(\bm{k})\right\rangle \! \right\rangle.
\label{eq:Kramers3}
\end{align}
It should be noted that this relation follows from the special property of the pseudo-time-reversal operator, i.e., Eq. (\ref{eq:Theta}).
From Eq. (\ref{eq:Kramers3}) for the TRIM $(k=\Lambda)$, we find that the two vectors $\bm{\psi}(\bm{\Lambda})$ and $\Theta'\bm{\psi}(\bm{\Lambda})$ are orthogonal, 
\begin{align}
\left\langle \! \left\langle \bm{\psi}(\bm{\Lambda}), \Theta'\bm{\psi}(\bm{\Lambda})\right\rangle \! \right\rangle=0.
\label{eq:Kramers4}
\end{align}
Therefore, the ``Kramers pairs'' of bosons $\bm{\psi}(\bm{\Lambda})$ and $\Theta'\bm{\psi}(\bm{\Lambda})$ can be defined 
under pseudo-time-reversal symmetry described by Eqs. (\ref{eq:Theta}) and (\ref{eq:Commutation}).

\subsection{The form of the Hamiltonian with the pseudo-time-reversal symmetry\label{Ham_PTRS}}

We consider a magnetically ordered system on a lattice which can be divided into two magnetic sublattices. All the spins in one magnetic sublattice point upward, while all the spins in the other magnetic sublattice point in the opposite direction. For convenience, we refer to the former the up spins and the latter the down spins.
For such a system, the magnon creation operator $\bm{\beta}^{\dagger}(\bm{k})$ (See Eq.~(\ref{eq:op_phi})) can generally be written as
\begin{align}
\bm{\beta}^{\dagger}(\bm{k})=[\bm{b}_{\uparrow}^{\dagger}(\bm{k}),\bm{b}_{\downarrow}^{\dagger}(\bm{k})],
\end{align}
where the creation operators of magnons originating from the up spins $\bm{b}_{\uparrow}^{\dagger}(\bm{k})$ and the down spins $\bm{b}_{\downarrow}^{\dagger}(\bm{k})$ are given by
\begin{align}
&\bm{b}_{\uparrow}^{\dagger}(\bm{k})=[b_{\uparrow,1}^{\dagger}(\bm{k}),\cdots,b_{\uparrow,N}^{\dagger}(\bm{k})],  \nonumber \\
&\bm{b}_{\downarrow}^{\dagger}(\bm{k})=[b_{\downarrow,1}^{\dagger}(\bm{k}),\cdots,b_{\downarrow,N}^{\dagger}(\bm{k})].
\label{eq:opvec}
\end{align}
Here, $N$ is the number of the sublattices in a unit cell and the operator $ b^\dagger_{\uparrow, i} ({\bm k}) $ ($b^\dagger_{\downarrow,i} ({\bm k}) $) creates a magnon originating from the spin pointing upward (downward) at site $i$. 
Now we introduce a concrete expression of the pseudo-time-reversal operator:
\begin{align}
\Theta'=(\sigma_z \otimes {\rm i} \sigma_{y} \otimes 1_{N})K.
\label{eq:pseudoTR}
\end{align}
The part $\sigma_z$ acts on the particle-hole space, while ${\rm i}\, \sigma_y$ interchanges the up and down spins with an extra sign. 
With this $\Theta'$, the most general Hamiltonian satisfying Eq.~(\ref{eq:Commutation}) takes the form: 
\begin{align}
H(\bm{k})=
\left(
\begin{array}{cccc}
h_{1}(\bm{k}) &h_{2}(\bm{k}) &\Delta_{2}(\bm{k}) &\Delta_{1}(\bm{k}) \\
h_{2}^{\dagger}(\bm{k}) &h_{1}^{*}(-\bm{k}) &\Delta_{1}^{*}(-\bm{k})&-\Delta_{2}^{\dagger}(\bm{k}) \\
\Delta_{2}^{\dagger}(\bm{k}) &\Delta_{1}^{*}(-\bm{k}) &h_{1}^{*}(-\bm{k}) &h_{2}^{*}(-\bm{k}) \\
\Delta_{1}(\bm{k}) &-\Delta_{2}(\bm{k}) &h_{2}^{T}(-\bm{k}) &h_{1}(\bm{k}) \\
\end{array} 
\right),
\label{eq:H}
\end{align}
where $h_i(\bm{k})$ and $\Delta_i(\bm{k})$ for $i=1, 2$ are $N \times N$ matrices and satisfy $h_{1}^{\dagger}(\bm{k})=h_{1}(\bm{k}),\Delta_{1}^{\dagger}(\bm{k})=\Delta_{1}(\bm{k}),h_{2}^{T}(\bm{k})=-h_{2}(-\bm{k}),$ and $\Delta_{2}^{T}(\bm{k})=\Delta_{2}(-\bm{k})$~\cite{Kondo19a,Kondo19b}. 

We now compare the pseudo-time-reversal operator with the time-reversal operator and see the similarities and differences between them.
Here we refer to the operator which interchanges the up and down spins without extra sign as the time-reversal operator.
This operator is defined as $\Theta=(\sigma_z \otimes \sigma_{x} \otimes 1_{N})K$.
We note that since this satisfies $\Theta^{2}=1$, the time-reversal symmetry does not ensure the existence of ``Kramers pairs" of magnons.
If the system is symmetric under interchanging the up and down spins, the Hamiltonian satisfies the time-reversal symmetry: $\Sigma_{z}H(-\bm{k})\Theta-\Theta\Sigma_{z}H(\bm{k})=0$.
The most general Hamiltonian satisfying the time-reversal-symmetry takes the form: 
\begin{align}
H(\bm{k})=
\left(
\begin{array}{cccc}
\bar{h}_{1}(\bm{k}) &\bar{h}_{2}(\bm{k}) &\bar{\Delta}_{2}(\bm{k}) &\bar{\Delta}_{1}(\bm{k}) \\
\bar{h}_{2}^{\dagger}(\bm{k}) &\bar{h}_{1}^{*}(-\bm{k}) &\bar{\Delta}_{1}^{*}(-\bm{k})&\bar{\Delta}_{2}^{\dagger}(\bm{k}) \\
\bar{\Delta}_{2}^{\dagger}(\bm{k}) &\bar{\Delta}_{1}^{*}(-\bm{k}) &\bar{h}_{1}^{*}(-\bm{k}) &\bar{h}_{2}^{*}(-\bm{k}) \\
\bar{\Delta}_{1}(\bm{k}) &\bar{\Delta}_{2}(\bm{k}) &\bar{h}_{2}^{T}(-\bm{k}) &\bar{h}_{1}(\bm{k}) \\
\end{array} 
\right),\label{eq:H2}
\end{align}
where $\bar{h}_i(\bm{k})$ and $\bar{\Delta}_i(\bm{k})$ for $i=1, 2$ are $N \times N$ matrices and satisfy $\bar{h}_{1}^{\dagger}(\bm{k})=\bar{h}_{1}(\bm{k}),\bar{\Delta}_{1}^{\dagger}(\bm{k})=\bar{\Delta}_{1}(\bm{k}),\bar{h}_{2}^{T}(\bm{k})=\bar{h}_{2}(-\bm{k}),$ and $\bar{\Delta}_{2}^{T}(\bm{k})=\bar{\Delta}_{2}(-\bm{k})$. 
We note that the only difference occurs in the spin-non-conserving terms: $\bar{h}_{2}(\bm{k})$ and $\bar{\Delta}_{2}(\bm{k})$.
The matrix $\bar{h}_{2}(\bm{k})$ satisfies the condition different from that of $h_{2}(\bm{k})$.
The $(2,4)$ and $(4,2)$ components of Eqs.~(\ref{eq:H}) and (\ref{eq:H2}) differ in their signs.
This means that the time-reversal and the pseudo-time-reversal symmetries are equivalent in a system without spin-non-conserving terms.
In such a case, the time-reversal symmetry ensures the existence of Kramers pairs.
Indeed, the magnon spin Hall systems proposed in the previous studies~\cite{Zyuzin16,Nakata17b} fall into this category.

\subsection{Periodic table for non-Hermitian topological phases\label{NHPT}}
As discussed in Sec.~\ref{boson_BdG}, bosonic BdG systems have non-Hermitian property, so that the topological characterization of Hermitian systems cannot be applied to the magnon systems.
Here we review the topological classification of non-Hermitian systems, according to Ref.~\cite{Kawabata19}.
At the end of this section, we discuss several examples of magnon topological phases and their classes.

The fundamental topological classification is based on the set of internal (non-spatial) symmetries: time-reversal symmetry (TRS), particle-hole symmetry (PHS), and chiral symmetry (CS), which is referred to as AZ symmetry.
In addition, the non-Hermitian Hamiltonian ${\tilde H}(\bm{k})$ does not satisfy ${\tilde H}^{*}(\bm{k}) = {\tilde H}^{T}(\bm{k})$, which gives rise to extra internal symmetry other than AZ symmetry, AZ$^{\dagger}$ symmetry. 
The AZ and AZ$^{\dagger}$ symmetries for gapped non-Hermitian systems are summarized in Tab.~\ref{table:AZ}.
TRS and PHS impose the following conditions on Hamiltonian $\tilde{H}(\bm{k})$:
\begin{align}
&\mathcal{T}_{+}^{-1}\tilde{H}^{*}(\bm{k})\mathcal{T}_{+}=\tilde{H}(-\bm{k}), \label{eq:Tplus1}\\
&\mathcal{T}_{+}\mathcal{T}^{*}_{+}=\pm 1\label{eq:Tplus2},
\end{align}
\begin{align}
&\mathcal{C}_{-}^{-1}\tilde{H}^{T}(\bm{k})\mathcal{C}_{-}=-\tilde{H}(-\bm{k}), \label{eq:Cminus1}\\
&\mathcal{C}_{-}\mathcal{C}_{-}^{*}=\pm 1 \label{eq:Cminus2}.
\end{align}
On the other hand, TRS$^\dagger$ and PHS$^\dagger$ impose the following conditions on $ {\tilde H}({\bm k}) $:
\begin{align}
&\mathcal{C}_{+}^{-1}\tilde{H}^{T}(\bm{k})\mathcal{C}_{+}=\tilde{H}(-\bm{k}), \label{eq:Cplus1}\\
&\mathcal{C}_{+}\mathcal{C}_{+}^{*}=\pm 1\label{eq:Cplus2},
\end{align}
\begin{align}
&\mathcal{T}_{-}^{-1}\tilde{H}^{*}(\bm{k})\mathcal{T}_{-}=-\tilde{H}(-\bm{k}), \label{eq:Tminus1}\\
&\mathcal{T}_{-}\mathcal{T}^{*}_{-}=\pm 1\label{eq:Tminus2},
\end{align}
where $\mathcal{T}_{\pm}$ and $\mathcal{C}_{\pm}$ are unitary matrices.
The chiral symmetry CS is a combination of TRS and PHS (or TRS$^{\dagger}$ and PHS$^{\dagger}$):
\begin{align}
&\Gamma^{-1}\tilde{H}^{\dagger}(\bm{k})\Gamma=-\tilde{H}(\bm{k}), \label{eq:Gamma1}\\
&\Gamma = \mathcal{T}_{\pm} \mathcal{C}_{\mp}\label{eq:Gamma2}
\end{align}

Pseudo-Hermiticity which is a generalization of Hermiticity plays an important role in non-Hermitian systems~\cite{Mostafazadeh02a,Mostafazadeh02b,Mostafazadeh10,Lee19,Ghatak19,Zhou19,Lieu18}. A Hamiltonian $ {\tilde H}({\bm k}) $ is said to be pseudo-Hermitian if it satisfies 
\begin{align}
&\eta \tilde{H}^{\dagger}(\bm{k}) \eta^{-1}=\tilde{H}(\bm{k}),\label{eq:pH} \\
&\eta^{2}=1,
\end{align}
where $ \eta $ is a unitary and Hermitian matrix~\cite{Fleury15}. The presence of the operator $ \eta $ commuting or anticommuting with symmetry operators plays a crucial role in the classification of topological phases of non-Hermitian systems. The results obtained in Ref.~\cite{Kawabata19} are summarized in Tab.~\ref{table:PT1} and Tab.~\ref{table:PT2}. 

The effective Hamiltonian matrix of a bosonic BdG Hamiltonian, $ {\tilde H}({\bm k}) = \Sigma_z H ({\bm k}) $, is pseudo-Hermitian with respect to $ \eta = \Sigma_z $. This implies the reality of the spectrum of $ {\tilde H}({\bm k}) $ when $ H({\bm k}) $ is positive definite (see Appendix A for details). 
The bosonic BdG systems always respect PHS~(\ref{eq:Cminus1}) with ${\cal C}_{-}=\Sigma_{y}$ as implied by the statement (ii) in Sec.~\ref{boson_BdG}.
We note however that one should reconstruct the topological classification when the virtual "Fermi level" we consider is in an energy gap away from zero energy.
In this case, since this choice of ``Fermi level'' does not respect PHS, the topological classification of the bosonic BdG systems obeys that without PHS.

Let us discuss examples of bosonic topological phases and their classification. The 2D and 3D magnon systems we consider later have the pseudo-time-reversal symmetry with $\Theta'^{2}=\mathcal{T}_{+}\mathcal{T}^{*}_{+} =-1$.
The pseudo-time-reversal operator commutes with $\eta$, and hence our magnon systems in 2D/3D are categorized as class AII with $\eta_{+}$ whose entries are ${\mathbb Z}_{2}\oplus{\mathbb Z}_{2}$, according to Tab.~\ref{table:PT2}.
However, since the original Hamiltonian $H(\bm{k})$ of them is positive definite, ${\mathbb Z}_{2}\oplus{\mathbb Z}_{2}$ topological invariant reduces to the single ${\mathbb Z}_{2}$ index.

As other examples of the topological phases of bosonic BdG systems, 2D magnon thermal Hall system with dipolar interaction in Ref.~\cite{Matsumoto14},  triplonic analog of Su-Schrieffer-Heeger model in Ref.~\cite{Nawa19}, and triplonic analog of spin Hall insulator in Ref.~\cite{Joshi19} belong to class A with $\eta$, class BDI with $\eta_{++}$, and class AII with $\eta_{+}$ respectively.
For the same reason as in our magnon systems, the 2D magnon thermal Hall system in Ref.~\cite{Matsumoto14} is characterized by the single Chern number whereas Tab. \ref{table:PT1} indicates ${\mathbb Z}\oplus{\mathbb Z}$ invariant.

\begin{table}[!h]
\caption{AZ and AZ$^{\dagger}$ classes for non-Hermitian Hamiltonians.
TRS, PHS, TRS$^{\dagger}$, PHS$^{\dagger}$, and CS are defined by Eq.~(\ref{eq:Tplus1}) with Eq.~(\ref{eq:Tplus2}), Eq.~(\ref{eq:Cminus1}) with Eq.~(\ref{eq:Cminus2}), Eq.~(\ref{eq:Cplus1}) with Eq.~(\ref{eq:Cplus2}), Eq.~(\ref{eq:Tminus1}) with Eq.~(\ref{eq:Tminus2}), and Eq.~(\ref{eq:Gamma1}) with Eq.~(\ref{eq:Gamma2}), respectively. 
The absence of symmetries is denoted by ``$0$''.
The presence of the symmetries $ {\cal U} = {\cal T}_{\pm}, {\cal C}_{\pm} $ is denoted by either $ +1 $ or $ -1 $, depending on whether $ {\cal U} {\cal U}^{*}=+1 $ or $ -1 $. In the last column, the presence (absence) of chiral symmetry (CS) is denoted by $ 1 $ ($ 0 $). 
}
\begin{center}
{\tabcolsep=1mm
  \begin{tabular}{lllllllllllll}
    \hline
    \hline
    Symmetry   &   & TRS & PHS  & TRS$^{\dagger}$ & PHS$^{\dagger}$  & CS &  \\
    class   &   & $(\mathcal{T}_{+})$ & $(\mathcal{C}_{-})$  & $(\mathcal{C}_{+})$ & $(\mathcal{T}_{-})$  & $(\Gamma)$ &  \\   \hline
    Complex AZ  & A & 0& 0 & 0& 0 & 0   \\ 
       & AIII & 0& 0 & 0& 0 & 1    \\ 
    Real AZ  & AI & $+1$& 0 & 0& 0 & 0    \\ 
       & BDI & $+1$& $+1$ & 0& 0 & 1    \\
       & D & 0& $+1$ & 0& 0 & 0        \\
       & DIII & $-1$& $+1$ & 0& 0 & 1    \\
       & AII & $-1$& 0 & 0& 0 & 0    \\
       & CII & $-1$& $-1$ & 0& 0 & 1    \\
       & C & 0& $-1$ & 0& 0 & 0    \\
       & CI & $+1$& $-1$ & 0& 0 & 1    \\ 
    Real AZ$^{\dagger}$  & AI$^{\dagger}$ & 0& 0 & $+1$& 0 & 0    \\ 
       & BDI$^{\dagger}$ & 0& 0 & $+1$& $+1$ & 1    \\
       & D$^{\dagger}$ & 0& 0  & 0& $+1$ & 0        \\
       & DIII$^{\dagger}$ & 0& 0  & $-1$& $+1$ & 1    \\
       & AII$^{\dagger}$ & 0& 0  & $-1$& 0 & 0    \\
       & CII$^{\dagger}$ & 0& 0  & $-1$& $-1$ & 1    \\
       & C$^{\dagger}$ & 0& 0  & 0& $-1$ & 0    \\
       & CI$^{\dagger}$ & 0& 0  & $+1$& $-1$ & 1    \\  \hline   \hline
  \end{tabular}
}
\end{center}
\label{table:AZ}
\end{table}
\noindent

\begin{table}[!h]
\caption{The periodic table for non-Hermitian topological phases in the complex AZ symmetry class with the pseudo-Hermiticity (pH). 
We only show the cases with a real line gap which corresponds to the conventional energy gap in Hermitian systems.
The subscript of $\eta_{+}$ ($\eta_{-}$) denotes the commutation (anticommutation) relation to CS, i.e., $\eta_{+}\Gamma=\Gamma\eta_{+}$ ($\eta_{-}\Gamma=-\Gamma\eta_{-}$).
}
\begin{center}
{\tabcolsep=1mm
  \begin{tabular}{lccccccccccccc}
    \hline
    \hline
    pH &AZ class      & $d=0$  & $d=1$ & $d=2$  & $d=3$ &  \\
    \hline
    $\eta$         &A          &   $\mathbb{Z}\oplus\mathbb{Z}$  & 0                     & $\mathbb{Z}\oplus\mathbb{Z}$& 0                     &   \\ 
    $\eta_{+}$ &AIII      &   0                      & $\mathbb{Z}\oplus\mathbb{Z}$ & 0                     & $\mathbb{Z}\oplus\mathbb{Z}$ &    \\ 
     $\eta_{-}$ &AIII     &   $\mathbb{Z}$  & 0                     & $\mathbb{Z}$& 0                     &   \\  \hline   \hline
  \end{tabular}
}
\end{center}
\label{table:PT1}
\end{table}
\noindent

\begin{table}[!h]
\caption{The periodic table for non-Hermitian topological phases  with a real line gap in the real AZ symmetry class with pseudo-Hermiticity.  
The plus (minus) sign of the subscript of $\eta$ denotes the commutation (anticommutation) relation to TRS and/or PHS.
For example, $\eta_{+-}$ in class BDI satisfies the following commutation and anticommutation relations: $\eta_{+-}\mathcal{T}=\mathcal{T}\eta_{+-}$ and $\eta_{+-}\mathcal{C}=-\mathcal{C}\eta_{+-}$.
}
\begin{center}
{\tabcolsep=1mm
  \begin{tabular}{lccccccccccccc}
    \hline
    \hline
    pH &AZ class      & $d=0$  & $d=1$ & $d=2$  & $d=3$ &  \\
    \hline
    $\eta_{+}$    &AI          &   $\mathbb{Z}\oplus\mathbb{Z}$                  & 0                                     & 0                                   & 0             &   \\ 
    $\eta_{++}$ &BDI      &   $\mathbb{Z}_{2}\oplus\mathbb{Z}_{2}$    & $\mathbb{Z}\oplus\mathbb{Z}$     & 0                                   & 0 &    \\ 
    $\eta_{+}$   &D     &   $\mathbb{Z}_{2}\oplus\mathbb{Z}_{2}$   & $\mathbb{Z}_{2}\oplus\mathbb{Z}_{2}$   & $\mathbb{Z}\oplus\mathbb{Z}$& 0                     &   \\ 
    $\eta_{++}$ &DIII     &   0   & $\mathbb{Z}_{2}\oplus\mathbb{Z}_{2}$   & $\mathbb{Z}_{2}\oplus\mathbb{Z}_{2}$& $\mathbb{Z}\oplus\mathbb{Z}$  &   \\      
   $\eta_{+}$ &AII     &   $2\mathbb{Z}\oplus2\mathbb{Z}$   & 0   & $\mathbb{Z}_{2}\oplus\mathbb{Z}_{2}$& $\mathbb{Z}_{2}\oplus\mathbb{Z}_{2}$ &   \\ 
   $\eta_{++}$ &CII     &   0   & $2\mathbb{Z}\oplus2\mathbb{Z}$   & 0& $\mathbb{Z}_{2}\oplus\mathbb{Z}_{2}$ &   \\ 
   $\eta_{+}$ &C     &   0   & 0   & $2\mathbb{Z}\oplus2\mathbb{Z}$ & 0 &   \\
   $\eta_{++}$ &CI     &   0   & 0   & 0 & $2\mathbb{Z}\oplus2\mathbb{Z}$ &   \\
   $\eta_{+-}$ &BDI     &   $\mathbb{Z}$   & 0   & 0 & 0 &   \\
   $\eta_{-+}$ &DIII     &   $\mathbb{Z}_{2}$   & $\mathbb{Z}_{2}$   & $\mathbb{Z}$ & 0 &   \\ 
   $\eta_{+-}$ &CII     &   $2\mathbb{Z}$   & 0   & $\mathbb{Z}_{2}$ & $\mathbb{Z}_{2}$ &   \\
   $\eta_{-+}$ &CI     &   0   & 0   & $2\mathbb{Z}$ & 0 &   \\
   $\eta_{-}$ &AI     &   $\mathbb{Z}$  & 0   & $\mathbb{Z}$ & 0 &   \\ 
   $\eta_{--}$ &BDI     &   0  & $\mathbb{Z}$   & 0 & $\mathbb{Z}$ &   \\ 
   $\eta_{-}$ &D     &   $\mathbb{Z}$  & 0   & $\mathbb{Z}$ & 0 &   \\ 
   $\eta_{--}$ &DIII     &   0  & $\mathbb{Z}$   & 0 & $\mathbb{Z}$ &   \\ 
   $\eta_{-}$ &AII    &  $\mathbb{Z}$  & 0   & $\mathbb{Z}$ & 0 &   \\
   $\eta_{--}$ &CII     &   0  & $\mathbb{Z}$   & 0 & $\mathbb{Z}$ &   \\
   $\eta_{-}$ &C    &  $\mathbb{Z}$  & 0   & $\mathbb{Z}$ & 0 &   \\ 
   $\eta_{--}$ &CI     &   0  & $\mathbb{Z}$   & 0 & $\mathbb{Z}$ &   \\
   $\eta_{-+}$ &BDI     &   $\mathbb{Z}_{2}$  & $\mathbb{Z}_{2}$   & $\mathbb{Z}$ & 0 &   \\ 
   $\eta_{+-}$ &DIII     &   $2\mathbb{Z}$  & 0   & $\mathbb{Z}_{2}$ & $\mathbb{Z}_{2}$ &   \\
   $\eta_{-+}$ &CII     &   0  & 0   & $2\mathbb{Z}$ & 0 &   \\
   $\eta_{+-}$ &CI     &   $\mathbb{Z}$   & 0   & 0 & 0 &   \\ \hline   \hline
  \end{tabular}
}
\end{center}
\label{table:PT2}
\end{table}
\noindent

\section{Topological phases of magnon BdG systems in 2D and 3D\label{magnon_2D_3D}}
In this section, we review the recent studies on the magnonic analog of 2D and 3D topological insulators and their $\mathbb{Z}_{2}$ topological invariants~\cite{Zyuzin16,Kondo19a,Kondo19b}.
Theoretical studies on $\mathbb{Z}_{2}$ magnon systems have developed as follows.
As the first symmetry-protected topological phases of magnons, a magnon spin Hall system with spin conservation~\cite{Zyuzin16} was proposed theoretically (Sec.~\ref{m_SHE_spincons}).
Such a system can be regarded as two copies of magnon thermal Hall systems so that the combined system restores the conventional time-reversal symmetry for bosons.
Afterward, by extending the idea of time-reversal symmetry in bosonic systems, we introduced pseudo-time-reversal symmetry which restricts the form of the Hamiltonian as expressed by Eq.~(\ref{eq:H}).
Owing to the extension and the form of Eq.~(\ref{eq:H}), we constructed a model of magnon $\mathbb{Z}_{2}$ topological phases with anisotropic exchange interactions breaking spin conservation (Sec.~\ref{honeycomb})~\cite{Kondo19a}.
Moreover, we gained  new insight from the model without spin conservation, and then further extended the concept of the magnon $\mathbb{Z}_{2}$ topological phases to 3D systems (Sec.~\ref{Z2_3DMT})~\cite{Kondo19b}.
As in $\mathbb{Z}_{2}$ topological insulators of fermions in 3D, the interactions breaking spin conservation is necessary to realize 3D topological magnon systems.
In this review, we also present a candidate material realizing the magnon spin Hall system.
In the following, we refer to magnonic analogs of 2D and 3D topological insulators as magnon spin Hall systems and 3D topological magnon systems, respectively.
For a summary of this section, see Table 4. 

\begin{table}[!h]
\caption{The summary of theoretical studies on symmetry-protected topological magnon phases discussed in this section.
The first and second columns list the dimension and the presence or absence of spin conservation of the systems. 
The sections in which we review the systems are given in the third column.
The fourth column shows which topological invariants are used to characterize the systems.
}
\begin{center}
{\tabcolsep=3mm
  \begin{tabular}{cccc}
    \hline \hline
    Dimension & Spin conservation & Section & Topological invariant  \\ \hline
    2D &YES &\ref{m_SHE_spincons}, \ref{model_2D} & Eqs.~(\ref{eq:SpinChern}) and (\ref{eq:D})  \\ \hline
    2D &NO & \ref{honeycomb} & Eq.~(\ref{eq:D})  \\ \hline
    3D &NO &\ref{Z2_3DMT}  & Eqs.~(\ref{eq:topo_inv}) and (\ref{eq:topo_inverse}) \\
    \hline \hline
  \end{tabular}
}
\end{center}
\label{table:flow}
\end{table}

\subsection{Magnon spin Hall systems and ${\mathbb Z}_2$ topological invariant\label{m_SHE}}
In this part, we discuss the construction of magnon spin Hall systems and the correspondence between their edge states and the topological invariant.
In Sec.~\ref{m_SHE_spincons}, we review previous studies on magnon spin Hall systems with spin conservation.
Section~\ref{m_SHE_Z2} provides the definition of the ${\mathbb Z}_2$  topological invariant for magnon spin Hall systems. In Sec.~\ref{model_2D} and~\ref{honeycomb} we construct models exhibiting the magnon spin Hall effect with and without spin conservation, respectively. 
In both models, we demonstrate the validity of the  ${\mathbb Z}_2$ topological invariant and confirm the correspondence between ${\mathbb Z}_2$ index and the presence of gapless edge states.
In addition, we present a candidate material realizing the magnon spin Hall system with spin conservation in Sec.~\ref{model_2D}.

\subsubsection{Magnon spin Hall systems with spin conservation\label{m_SHE_spincons}}
\begin{figure}[!h]
\centering
  \includegraphics[width=0.6\columnwidth]{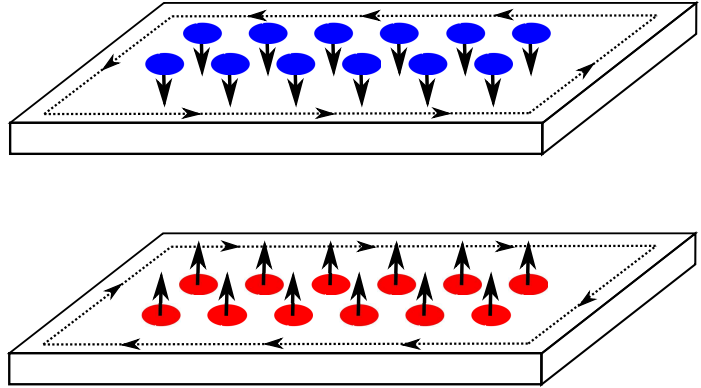}
\caption{A schematic picture  of a magnon spin Hall system. The number of up spins localized in one layer is the same as the number of down spins in the other layer. Magnon Hall current deriving from up and down spins (shown by the arrows along the edges) propagate in opposite directions.
}\label{fig:gairyaku}
\end{figure}

The theoretical models of magnon spin Hall systems are constructed~\cite{Zyuzin16,Nakata17b} by combining two magnon thermal Hall systems~\cite{Katsura10} with opposite magnetic moments.
The schematic picture of the magnon spin Hall system is shown in Fig.~\ref{fig:gairyaku}.
Hall current of magnons deriving from up and down spins propagate in opposite directions.
Since magnons from up and down spins convey down and up spins, respectively,
a nonzero spin current appears while the total energy current cancels out.
\noindent

\begin{figure}[!h]
\centering
  \includegraphics[width=0.6\columnwidth]{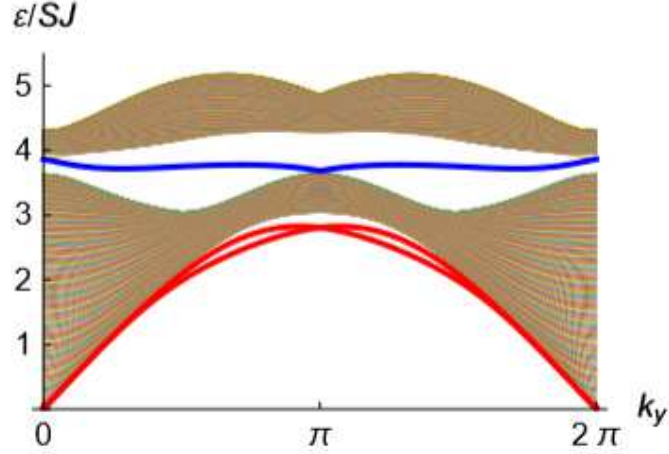}
\caption{The band structure of a strip of the magnon spin Hall system of honeycomb lattice bilayer antiferromagnets. The gapless helical edge state is shown in blue. This figure is taken from Ref.~\cite{Zyuzin16}. 
}\label{fig:Zyuzin16}
\end{figure}

The magnon spin Hall systems proposed in Ref.~\cite{Zyuzin16,Nakata17b} are the systems with spin conservation where time-reversal symmetry is identical to pseudo-time-reversal symmetry as mentioned in Sec.\ref{Ham_PTRS}.
Such systems can be divided into two independent magnon thermal Hall systems with up and down spins.
In this case with energy gap, each separated band can be characterized by the spin Chern number~\cite{Sheng05,Sheng06b} which is defined as the difference of the Chern numbers of up-spin (${\rm Ch}_{\uparrow}$) and down-spin part (${\rm Ch}_{\downarrow}$):
\begin{align}
({\rm Spin\hspace{1mm} Chern\hspace{1mm} number})=\frac{1}{2}\left({\rm Ch}_{\uparrow}-{\rm Ch}_{\downarrow}\right),
\label{eq:SpinChern}
\end{align}
while the conventional Chern number $({\rm Ch}_{\uparrow} + {\rm Ch}_{\downarrow})/2$ is zero due to (pseudo-)time-reversal symmetry.
Figure~\ref{fig:Zyuzin16} shows the magnon band structure in Ref.~\cite{Zyuzin16} with nonzero spin Chern number. The gapless helical edge state characterized by the nontrivial topological invariant contribute to the magnon spin Hall effect, resulting in the pure spin current.

\subsubsection{${\mathbb Z}_2$ topological invariant for magnon spin Hall systems\label{m_SHE_Z2}}
Here we discuss ${\mathbb Z}_2$ topological invariant for magnon spin Hall systems with/without spin conservation and the correspondence between the topological invariant and helical edge states.

Here we shall introduce a definition of the ${\mathbb Z}_2$ topological invariant for bosonic systems with the pseudo-time-reversal symmetry. 
For fermionic systems, there are various definitions of the ${\mathbb Z}_2$ invariant~\cite{Fu06, Fukui07, Moore07, Fukui08, Qi08, Roy09, Wang10, Loring10, Fulga12, Sbierski14, Loring15, Katsura16, Akagi17, Katsura18}. 
Here we follow the approach developed by Fu and Kane~\cite{Fu06}. 

\begin{figure}[!h]
\centering
  \includegraphics[width=0.8\columnwidth]{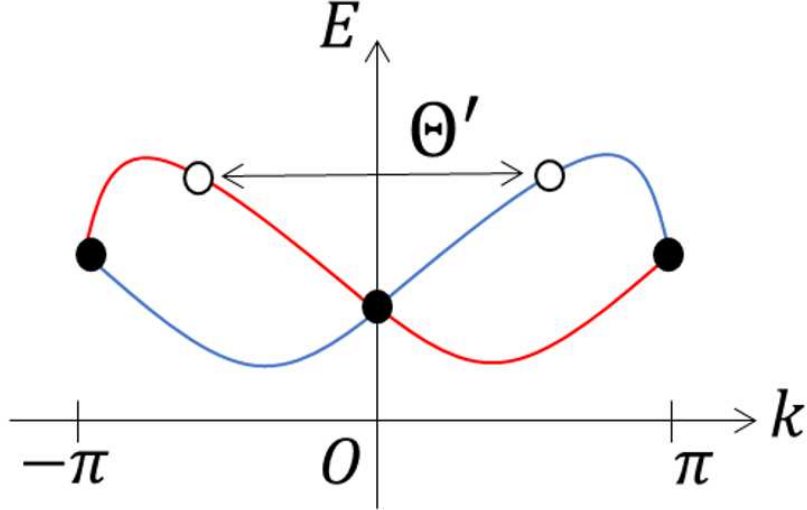}
\caption{Schematic picture of a Kramers pair of bands.
The energy spectra for the $n$th Kramers pair $E_{n}(\bm{k})$ and $E_{n}(-\bm{k})$ are shown in red and blue, respectively.
They degenerate at TRIM: $k=0,\pi$ which is indicated by black dots.
The white dots indicate generic points in the bands which are related by the pseudo-time-reversal operator.
}\label{fig:Kramers}
\end{figure}

Let ${\bm \Psi}_{n,1,+}({\bm k})$ ($n=1,\cdots,{\mathscr N}/2$) be an eigenvector of $\Sigma_z H({\bm k})$ with eigenvalue $E_n({\bm k}) \ge 0$, i.e., a particle wavefunction. 
As explained in Sec.~\ref{Kramers_boson}, ${\bm \Psi}_{n,2,+}({\bm k}):=-\Theta' {\bm \Psi}_{n,1,+} (-{\bm k})$ is also an eigenvector of $\Sigma_z H({\bm k})$ with eigenvalue $E_n(-{\bm k})$, and forms the $n$th Kramers pair with $\Psi_{n,1,+}$. 
Figure~\ref{fig:Kramers} shows a schematic picture of bosonic energy bands with the Kramers pair and the Kramers degeneracy at a TRIM. 
The particle-hole conjugates ${\bm \Psi}_{n,l,-}({\bm k}) = \Sigma_x K {\bm \Psi}_{n,l,+}(-{\bm k}) ,(l=1,2)$ are the eigenvectors of $\Sigma_z H({\bm k})$ with eigenvalue $-E_n((-1)^l {\bm k})$ as described by (ii) of Sec.~\ref{boson_BdG}. It follows from the para-unitarity that the wavefunctions obey  $\langle \! \langle {\bm \Psi}_{n,l,\sigma} ({\bm k}), {\bm \Psi}_{m,l',\sigma'} ({\bm k}) \rangle \! \rangle = \sigma \delta_{n,m} \delta_{l,l'} \delta_{\sigma,\sigma'} (\sigma=\pm)$.

The Berry connection and curvature for the $n$th Kramers pair of particle- (hole-) bands are defined as
\begin{align}
&\bm{A}_{n, \sigma}(\bm{k})=\sum_{l=1,2}\bm{A}_{n,l,\sigma}(\bm{k}) \label{eq:BdGBerryconnection},
\\
&\bm{\Omega}_{n,\sigma}(\bm{k})=\sum_{l=1,2}\bm{\Omega}_{n,l,\sigma}(\bm{k}) \label{eq:BdGBerrycurvature}, 
\end{align}
where 
\begin{align}
&\bm{A}_{n,l,\sigma}(\bm{k})
={\rm i}\, \sigma \left\langle \! \left\langle \bm{\Psi}_{n,l,\sigma}(\bm{k}),\nabla_{\bm{k}}\bm{\Psi}_{n,l,\sigma}(\bm{k})\right\rangle \! \right\rangle,
\\
&\bm{\Omega}_{n,l,\sigma}(\bm{k})=\nabla_{\bm{k}} \times \bm{A}_{n,l,\sigma}(\bm{k}) .
\end{align} 
The Berry connections of the particle bands and those of the hole bands are related to each other via ${\bm A}_{n,1,+}({\bm k})={\bm A}_{n,2,-}({\bm k})$ and ${\bm A}_{n,2,+}({\bm k})={\bm A}_{n,1,-}({\bm k})$, yielding ${\bm A}_{n,\sigma}({\bm k})={\bm A}_{n,-\sigma}({\bm k})$ and $\bm{\Omega}_{n,\sigma}({\bm k})=\bm{\Omega}_{n,-\sigma}({\bm k})$. 

Using $\bm{A}_{n,\sigma}$ and $\bm{\Omega}_{n,\sigma}$, the ${\mathbb Z}_2$ index of the $n$th Kramers pair of bands for magnon spin Hall systems is defined as 
\begin{align}
D_{n,\sigma} \! := \frac{1}{2\pi} \! \left[ \oint_{\partial {\rm EBZ}}d\bm{k} \cdot \bm{A}_{n,\sigma}(\bm{k})-\int_{\rm EBZ}  d^2k \Omega_{n,\sigma}^{z}(\bm{k}) \right]\hspace{1mm}{\rm mod}\hspace{1mm}2, 
\label{eq:D}
\end{align}
where EBZ and ${\partial {\rm EBZ}}$ stand for the effective Brillouin zone and its boundary, respectively. The EBZ related to the time-reversal-invariant band structures describes one-half of the Brillouin zone (e.g., see Fig. \ref{fig:kagome}(a)). 
Equation (\ref{eq:D}) is the main result of this section. 
Since the relation $D_{n, \sigma} = D_{n, -\sigma}$ holds as mentioned in Sec.~\ref{NHPT}, we drop the subscript $\sigma=\pm$ in the following.
We note in passing that the magnon Chern number $C_{n,l}$ ($l=1,2$) is given by $C_{n,l}=\frac{1}{2\pi}\int_{\rm BZ} d^2k \Omega_{n,l,+}(\bm{k})$.

\subsubsection{First model: kagome bilayer system\label{model_2D}}

\begin{figure}[!h]
\centering
  \includegraphics[width=0.90\columnwidth]{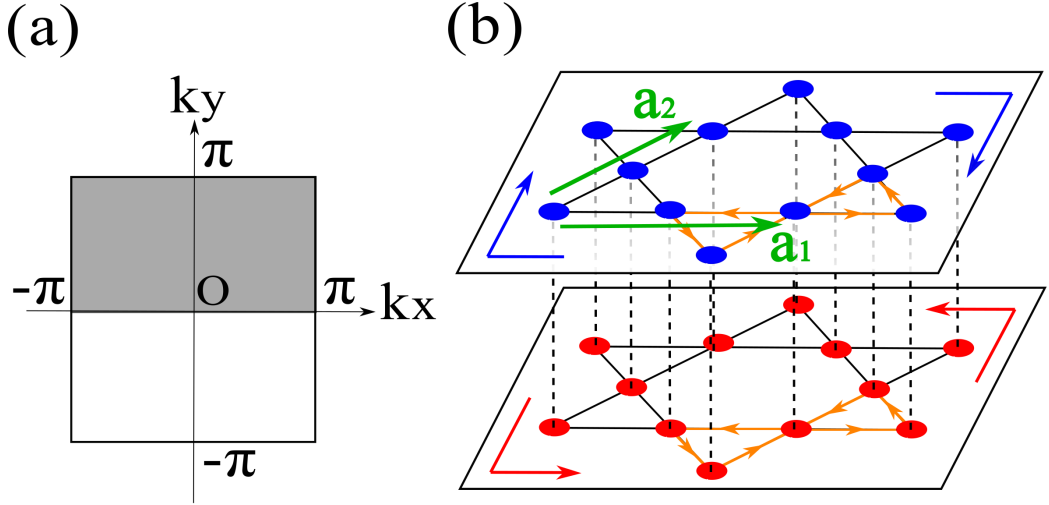}
\caption{(a) The Brillouin zone (BZ) and the effective Brillouin zone (EBZ) indicated by the shaded region. 
(b) The ``ferromagnetic" bilayer kagome system exhibiting the magnon spin Hall effect. 
The red and blue dots indicate up and down spins, respectively.
The vectors $\bm{a}_{1}$ and $\bm{a}_{2}$ are the primitive lattice vectors. 
The orange arrows on the triangular plaquette represent the sign convention for the DM vectors. 
Magnon edge states with opposite magnetic dipole moments propagate in opposite directions, as shown by the red and blue arrows.
Taking the primitive lattice vector as $\bm{a}_{1} = (1, 0)$ and $\bm{a}_{2} = (0, 1)$, we deform the shape of BZ of the kagome lattice into that of the square lattice as shown in (a).
The figures are taken from Ref.~\cite{Kondo19a}.
}\label{fig:kagome}
\end{figure}
This section provides a model showing the magnon spin Hall effect with spin conservation. We also demonstrate the validity of the definition of the ${\mathbb Z}_{2}$ index for the model. In addition, we propose a candidate material of such a magnon spin Hall system at the end of this section.

Let us consider a ``ferromagnetic" bilayer kagome system without net-moment. 
Here, we assume that the spins on each layer are ferromagnetically ordered while the directions of spins on the two layers are opposed the each other via interlayer antiferromagnetic interaction (see Fig.~\ref{fig:kagome}(b)). 
The Hamiltonian is given by Eq.~(14) in Ref.~\cite{Kondo19a}.

\begin{figure}[!h]
  \centering
  \includegraphics[width=0.97\columnwidth]{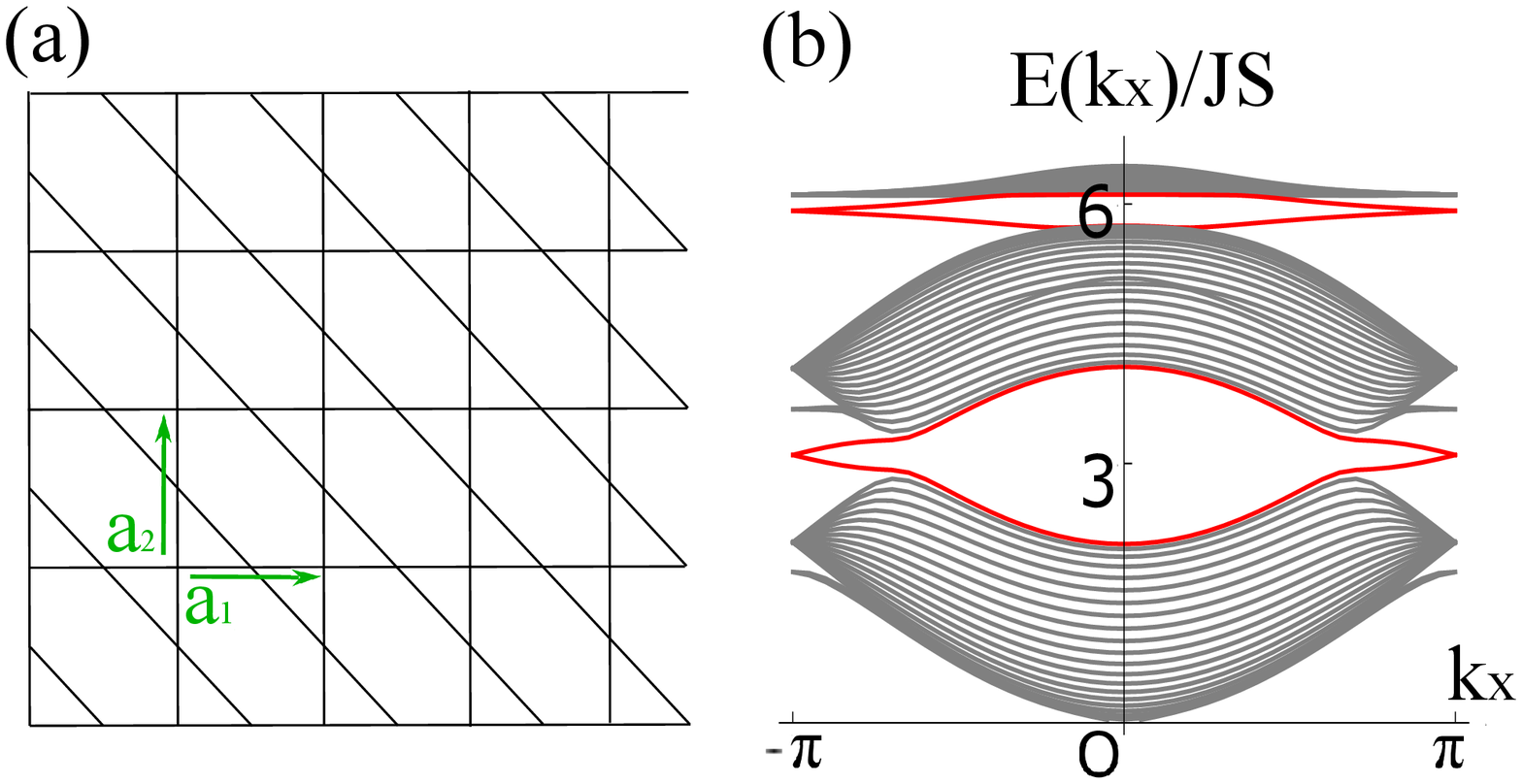}
\caption{(a) Cylindrical boundary conditions on the kagome lattice with $M$ unit cells in the vertical direction.
(b) Magnon spectrum of the bilayer kagome lattice under the cylindrical boundary condition with $M=20$ width for $D=0.1 J$, $J'=0.1J$, and $J>0$. 
Magnon helical edge states protected by pseudo-time-reversal symmetry shown in red occur in each energy gap.
The figures are taken from Ref.~\cite{Kondo19a}.}
    \label{fig:edge} 
\end{figure}

Figure~\ref{fig:edge}(b) shows the magnon spectrum in the bilayer kagome system with cylindrical boundary conditions (Fig. \ref{fig:edge}(a)). 
Each band is exactly degenerate not only at TRIMs but all $ k_x $. This is because, in addition to time-reversal symmetry, the Hamiltonian $ H({\bm k}) $ has a $ k $-dependent symmetry $ U({\bm k}) =1_{4}\otimes{\rm diag}[1, e^{{\rm i}k_{x}}, e^{{\rm i}k_{y}}]$ which acts on $ H({\bm k}) $ as $ U({\bm k})^{-1} H({\bm k}) U({\bm k}) = H(-{\bm k}) $.
The distinctive feature of the spectrum is the edge states, which traverse the energy gaps. 
Correspondingly, using Eq.~(\ref{eq:D}) and the numerical method by Ref.~\cite{Fukui07}, we obtain that the ${\mathbb Z}_2$ indices are 1, 0, and 1 from the lowest band to the highest band, i.e., 
$D_{1}=1$, $D_{2}=0$, and $D_{3}=1$. 
The indices remain the same by changing the parameters as long as the aforementioned magnetic order is stable. 

\begin{table}[!h]
\caption{
The relation between magnon Chern numbers and $\mathbb{Z}_2$ indices of particle bands.
Here, $C_{n,l}$ denotes the Chern number labeled by the band index $n$ and index of Kramers pair $l$, while $D_{n}$ is the $\mathbb{Z}_2$ topological invariant of the $n$th Kramers pair of bands for the bilayer kagome system\protect\footnotemark. 
Each Kramers pair with ${\mathbb Z}_2$ index unity consists of two bands with Chern numbers $+1$ and $-1$.}
\begin{center}
{\tabcolsep=5mm
  \begin{tabular}{cccc}
    \hline \hline
    $n$ & $C_{n,1}$ & $C_{n,2}$ & $D_{n}$ \\ \hline
    $1$ (top) & $+1$ & $-1$ & 1 \\ \hline
    $2$ (middle) & 0 & 0 & 0 \\ \hline
    $3$ (bottom) & $-1$ & $+1$ & 1 \\
    \hline \hline
  \end{tabular}
}
\end{center}
\label{table:chern}
\end{table}
\footnotetext{The Chern numbers ($\mathbb{Z}_2$ indices) of the three hole bands are the opposite (same) to those of the corresponding particle bands.}
As is clear from Tab.~\ref{table:chern}, the nontrivial ${\mathbb Z}_2$ indices come from the pair of magnon Chern numbers, $+1$ and $-1$. 
In fact, owing to the spin conservation, we can regard the ${\mathbb Z}_2$ index as the spin Chern number of magnons $D_{n}=\frac{1}{2}(C_{n,1}-C_{n,2})$ (mod $2$), as in electronic systems with the conservation of $S_z$. 
Because of the pseudo-time-reversal symmetry, the total Chern number of each Kramers pair vanishes, i.e., $C_n=C_{n,1}+C_{n,2}=0$. 
Correspondingly, the system exhibits not thermal Hall effect but magnon spin Hall effect by pure spin current.

\begin{figure}[!h]
  \centering
  \includegraphics[width=0.97\columnwidth]{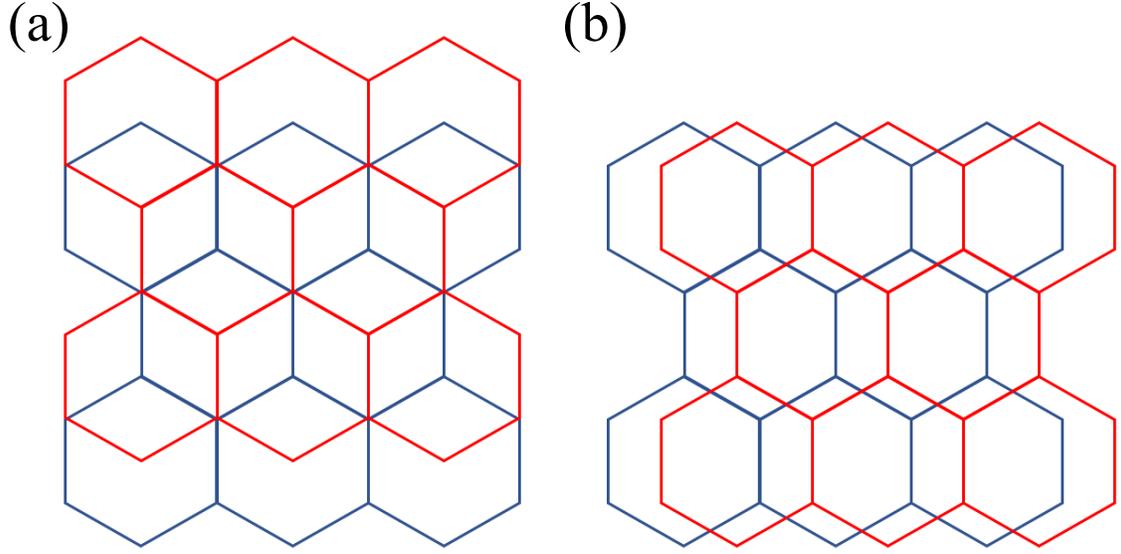}
\caption{(Color online) (a) Rhombohedral and (b) monoclinic structure of the honeycomb lattice. The first and second layers are shown in red and blue, respectively.
Magnetic compound CrI$_{3}$ with the former and the latter structures have interlayer ferromagnetic and antiferromagnetic interactions, respectively.}
\label{fig:structure}
\end{figure}

So far we have considered a specific example for concreteness. However, the realization of a magnon spin Hall system is not limited to such an example. In fact, there is a way to construct a system with desired properties in a more systematic manner. To illustrate this, let us consider a system consisting of two antiferromagnetically coupled ferromagnetic layers. For such a system, one can prove that the Berry connection and curvature perfectly coincide with those of the two independent single layer systems without interlayer coupling. 
This leads to the conclusion that the general bilayer ``ferromagnet" also exhibits the magnon spin Hall effect due to a nonzero spin Chern number (see Appendix \ref{bi_ferro} for details).
Thanks to the generalization, we have found that bilayer CrI$_{3}$ is a candidate material realizing the magnon spin Hall system~\cite{McGuire15,Huang17,Huang18,Chen18,Sivadas18,Soriano19}.
Magnetic compound CrI$_{3}$ is a layered honeycomb lattice material with the intralayer ferromagnetic and DM interaction.
In this material, the magnetic moments are carried by Cr$^{3+}$ ions with electronic configuration $3d^3$.  
The spin magnitude of each Cr$^{3+}$ ion is $S=3/2$.
Bulk CrI$_{3}$ has stacking structures called rhombohedral and monoclinic at low and high temperatures, respectively (see Fig.~\ref{fig:structure}).
Due to the difference of the structures, the interlayer interactions of the former and the latter are ferromagnetic and antiferromagnetic, respectively.
Recently it has been reported that the monoclinic structure can be realized at low temperatures in a thin film of CrI$_{3}$~\cite{Ubrig19}.
Since theoretical models of CrI$_{3}$ which do not have spin-conservation-breaking interaction give a reasonable explanation for the material~\cite{Chen18}, the above general construction method for magnon spin Hall systems, which is discussed in the case of spin conservation, is expected to be applied.
Thus, bilayer CrI$_{3}$ is a candidate material to investigate the magnon spin Hall effect.

\subsubsection{Second model: honeycomb bilayer system\label{honeycomb}}

\begin{figure}[!h]
  \centering
  \includegraphics[width=0.97\columnwidth]{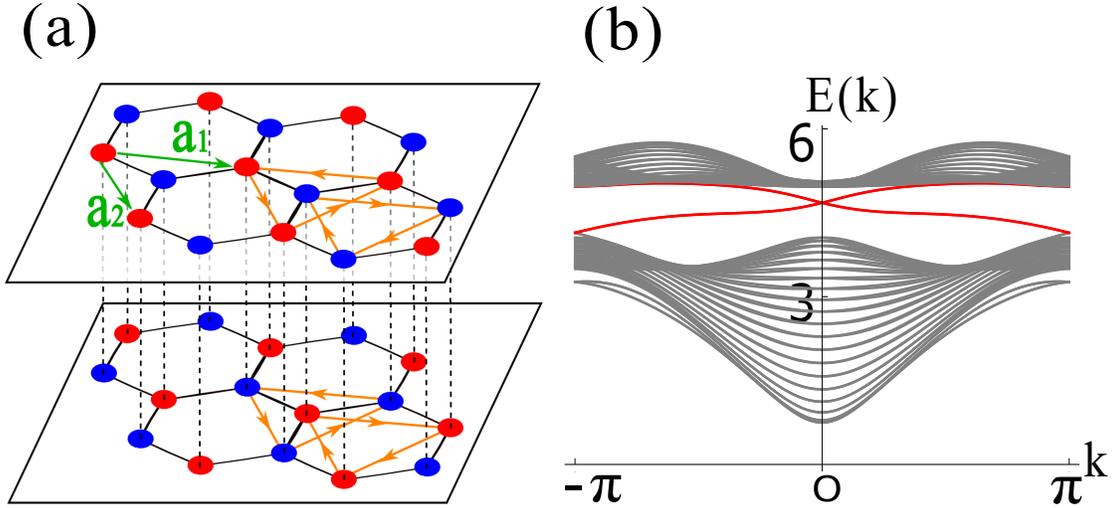}
\caption{(Color online) (a) The bilayer honeycomb system without spin conservation realizing a magnon ${\mathbb Z}_{2}$ topological phase. 
The two primitive lattice vectors are represented as $\bm{a}_{1}=(1,0)$ and $\bm{a}_{2}=(0,1)$. 
The red and blue dots indicate up and down spins, respectively, forming a perfect staggered spin configuration.
The orange arrows represent the sign convention of DM interaction $\xi_{ij}=+1 (=-\xi_{ji})$ for $i \to j$. 
(b) Magnon spectrum under the cylindrical boundary condition with zigzag edges of $M=20$ width for $J_x^{(1)} S=1.03$, $J_y^{(1)} S=0.97$, $J_z^{(1)} S=1.02$, $DS=0.2$, and $J' S=2$.
The magnon 
edge states are shown in red.
The figures are taken from Ref.~\cite{Kondo19a}.}
\label{fig:honeycomb}
\end{figure}
As a second example, we consider a bilayer antiferromagnetic honeycomb lattice system with a perfect staggered magnetic order, as shown in Fig.~\ref{fig:honeycomb}(a). The Hamiltonian is given by Eq. (16) in Ref.~\cite{Kondo19a}. In contrast to the previous example, this system does not preserve $S_z$, which is analogous to the Kane-Mele model with a finite Rashba spin-orbit coupling~\cite{Kane05b}. 
The spin-conservation-breaking term of the Hamiltonian comes from the anisotropic XYZ interaction.
Thus, the spin Chern number of magnons can no longer be used and the use of the original definition of the ${\mathbb Z}_{2}$ index is essential here.

Figure~\ref{fig:honeycomb}(b) shows the magnon spectrum of the bilayer honeycomb system under cylindrical boundary condition with zigzag edges.\footnote{The bilayer honeycomb system with armchair edges also exhibits similar helical edge states.} 
The helical edge states exist and cross the energy gap, as in the kagome bilayer system.
Applying Eq.~(\ref{eq:D}) to the system, we find the ${\mathbb Z}_2$ index of each magnon band $D_{n}=1$ for $n=1,2$, reflecting the presence of the helical edge states. 
Unlike the first example, the Berry connections and curvatures of this system cannot be reduced to those of the single layer system. 
The topological invariants remain unchanged under the change of parameters as long as the staggered magnetic order is stable. The helical edge states are expected to be responsible for 
the magnon spin Nernst effect studied in Ref.~\cite{Zyuzin16} if the XYZ term which breaks conservation of spin is almost isotropic.

\subsection{3D topological magnon systems\label{Z2_3DMT}}

In this section, we consider the generalization of the magnon spin Hall systems to 3D.
In Sec.~\ref{Z2_def_3D}, we define topological invariants for 3D topological magnon systems.
Sec.~\ref{model_3D} gives a model of the ${\mathbb Z}_{2}$ topological magnon systems on the diamond lattice.
By computing the topological invariants, we determine the phase diagram which includes the strong topological, weak topological, and trivial phases.
In Sec.~\ref{surf_Hall}, we also discuss a possible surface thermal Hall effect that is expected to occur in a heterostructure of a ferromagnet and a 3D topological magnon system.
This section is based on our paper~\cite{Kondo19b}.

\subsubsection{Topological invariants for 3D topological magnon systems\label{Z2_def_3D}}

By using the Berry connection Eq.~(\ref{eq:BdGBerryconnection}) and curvature Eq.~(\ref{eq:BdGBerrycurvature}) of bosons, we define the topological invariants for 3D topological magnonic (bosonic) systems as follows:
\begin{align}
&\nu_{i,0}^{n\sigma}:= \frac{1}{2\pi}  \left[ \oint_{\partial {\rm EBZ}_{i,0}} d\bm{k} \cdot \left[\bm{A}_{n\sigma}(\bm{k})\right]_{k_{i}=0} - \int_{{\rm EBZ}_{i,0}} dk_{j}dk_{k}  \left[\Omega_{n\sigma}^{i}(\bm{k})\right]_{k_{i}=0} \right]\hspace{1mm}{\rm mod}\hspace{1mm}2, \nonumber \\
&\nu_{i,\pi}^{n\sigma} := \frac{1}{2\pi}  \left[ \oint_{\partial {\rm EBZ}_{i,\pi}}  d\bm{k} \cdot \left[\bm{A}_{n\sigma}(\bm{k})\right]_{k_{i}=\pi} - \int_{{\rm EBZ}_{i,\pi}} dk_{j}dk_{k}  \left[\Omega_{n\sigma}^{i}(\bm{k})\right]_{k_{i}=\pi} \right]\hspace{1mm}{\rm mod}\hspace{1mm}2,  \label{eq:topo_inv}
\end{align}
where $n$ is a band index and $i=x$, $y$, and $z$.
Here, $j$ and $k$ represent two of $x,y$, and $z$ which are different from $i$.
The index $\sigma=\pm$ denotes the particle and hole space, respectively.
The definitions of ${\rm EBZ}_{x,0}$ and others are the same as those of electronic systems in Eqs.~(\ref{eq:e_Z2_0}) and (\ref{eq:e_Z2_pi}).
This topological invariants for 3D topological bosonic phases can also be easily calculated by using the numerical method of Ref.~\cite{Fukui07}.

As in the case of magnon spin Hall systems, the topological invariants of a particle and a hole have the same values: $\nu_{i,0}^{n+}=\nu_{i,0}^{n-},\nu_{i,\pi}^{n+}=\nu_{i,\pi}^{n-}$. 
In the following we write $\nu_{i,0(\pi)}^{n\pm}=\nu_{i,0(\pi)}^{n}$.
By introducing the virtual ``Fermi level" $\epsilon$ of bosons, the same correspondence holds between the summation of topological invariants over the bands below $\epsilon$ and the number of the surface states as that in 3D ${\mathbb Z}_{2}$ topological insulators for fermions~\cite{Fu07}. 
The summation counts the number of the surface states at the ``Fermi level" $\epsilon$ modulo 2.
As discussed for electron systems in Sec.~\ref{e_3DTI}, four of six topological index $\nu_{i,0(\pi)}$ are independent.
Here we define a set of independent topological indices $(\nu_{0}^{n};\nu_{x}^{n},\nu_{y}^{n},\nu_{z}^{n})$ as $\nu_{0}^{n}=\nu_{x,0}^{n}+\nu_{x,\pi}^{n}$ and $\nu_{i}^{n}=\nu_{i,\pi}^{n}$ $(i=x,y,z)$.
Following the discussion in Sec.~\ref{e_3DTI}, magnetic phases for $\nu_0=\sum_{n, (E_{n}(\bm{k}) \leq \epsilon)} \nu_0^n\hspace{3mm} (\rm{mod}\hspace{2mm} 2)=1$ ($\nu_0=0$ and at least one of $\nu_i = \sum_{n, (E_{n}(\bm{k}) \leq \epsilon)} \nu_{i}^{n}\hspace{3mm} (\rm{mod}\hspace{2mm} 2)$ $(i = x,y,z)$ taking nonzero) is named as the strong (weak) magnon topological phase.


In the following, we give an example of 3D topological magnon systems. We calculate the band structures of the system in a slab geometry, thereby confirming the correspondence of the topological invariants with the numbers and positions of surface Dirac cones.
As in the case of 2D systems, we consider a system in which the same number of up and down spins are localized.
The pseudo-time-reversal operator and the generic form of the Hamiltonian with the pseudo-time-reversal symmetry are given by the same form as Eqs.~(\ref{eq:pseudoTR}) and (\ref{eq:H}), respectively.

\subsubsection{Example: diamond lattice system\label{model_3D}}

\begin{figure}[!h]
\centering
  \includegraphics[width=9cm]{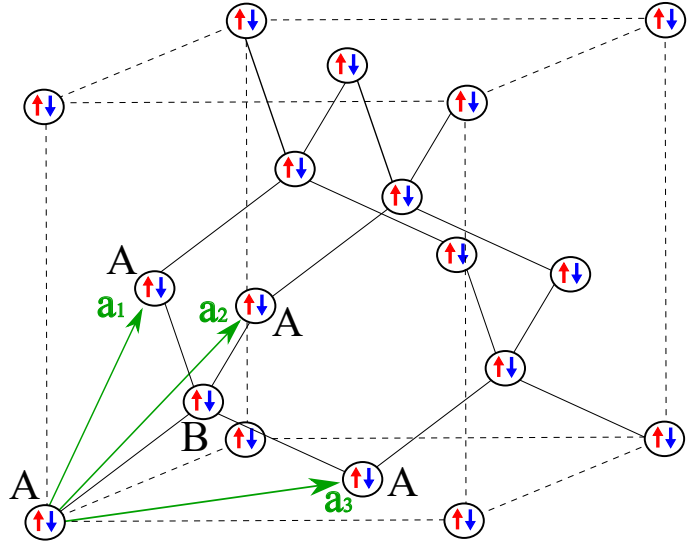}
\caption{Diamond lattice system where two spins are placed at each lattice site. 
Two sublattice indices are denoted by $A$ and $B$. The vectors $\bm{a}_{i}(i=1,2,3)$ are the lattice prime vectors. 
The figure is taken from Ref.~\cite{Kondo19b}.}\label{fig:cell_di}
\end{figure}

We provide an example of 3D topological magnon phases on the diamond lattice.
In this system depicted in Fig.~\ref{fig:cell_di}, we assume that two spins are localized at each site and aligned in the opposite direction to each other due to the antiferromagnetic interaction between them. 
The Hamiltonian of the system is given by Eq.~(17) in Ref.~\cite{Kondo19b}.
In this section, we take the spin magnitude $ S $ to be unity for simplicity.

The Hamiltonian of the system is written as
\begin{align}
H=H_{\rm DM}+H_{J'}+H_{J}+H_{\rm XY}+H_{\rm \Gamma}+H_{\kappa} .\label{eq:Ham_di}
\end{align}
Here, the first term $H_{\rm DM}$ is the next-nearest-neighbor DM interaction between spins which are aligned in the same direction.
The second term $H_{J'}$ is the antiferromagnetic interaction between two spins on the same site. 
The third term $H_{J}$ is the nearest-neighbor bond-dependent ferromagnetic interaction between spins pointing in the same direction.
The fourth and fifth terms $H_{\rm XY}$ and $H_{\rm \Gamma}$ are the next-nearest-neighbor anisotropic XY and ${\rm \Gamma}$ interactions between  spins aligned in opposite directions.
The sixth term $H_{\kappa}$ is the easy axis anisotropy.

By using the spin operators, the Hamiltonians of the interactions are expressed as follows:
\begin{align}
&H_{\rm DM}
=\sum_{\bm{R},s={\rm u,d}}
D_{1}^{z}(S_{s}^{x}(\bm{R},A)S_{s}^{y}(\bm{R}+\bm{a}_{1},A)-S_{s}^{y}(\bm{R},A)S_{s}^{x}(\bm{R}+\bm{a}_{1},A)) \nonumber \\
&\hspace{20mm}+D_{2}^{z}(S_{s}^{x}(\bm{R},A)S_{s}^{y}(\bm{R}+\bm{a}_{2},A)-S_{s}^{y}(\bm{R},A)S_{s}^{x}(\bm{R}+\bm{a}_{2},A)) \nonumber \\
&\hspace{20mm}+D_{3}^{z}(S_{s}^{x}(\bm{R},A)S_{s}^{y}(\bm{R}+\bm{a}_{3},A)-S_{s}^{y}(\bm{R},A)S_{s}^{x}(\bm{R}+\bm{a}_{3},A)) \nonumber \\
&\hspace{20mm}+D_{21}^{z}(S_{s}^{x}(\bm{R},A)S_{s}^{y}(\bm{R}+\bm{a}_{21},A)-S_{s}^{y}(\bm{R},A)S_{s}^{x}(\bm{R}+\bm{a}_{21},A)) \nonumber \\
&\hspace{20mm}+D_{31}^{z}(S_{s}^{x}(\bm{R},A)S_{s}^{y}(\bm{R}+\bm{a}_{31},A)-S_{s}^{y}(\bm{R},A)S_{s}^{x}(\bm{R}+\bm{a}_{31},A)) \nonumber \\
&\hspace{20mm}+D_{32}^{z}(S_{s}^{x}(\bm{R},A)S_{s}^{y}(\bm{R}+\bm{a}_{32},A)-S_{s}^{y}(\bm{R},A)S_{s}^{x}(\bm{R}+\bm{a}_{32},A)) \nonumber \\
&\hspace{20mm}-(A\leftrightarrow B) , \\
&H_{J'}=J'\sum_{i}\bm{S}_{i,{\rm u}}\cdot\bm{S}_{i,{\rm d}}, \\
&H_{J} 
=-\sum_{\bm{R},s={\rm u,d}}J_{0}\bm{S}_{s}(\bm{R},A)\cdot\bm{S}_{s}(\bm{R},B)+J_{1}\bm{S}_{s}(\bm{R},A)\cdot\bm{S}_{s}(\bm{R}+\bm{a}_{1},B) \nonumber \\
&\hspace{20mm}+J_{2}\bm{S}_{s}(\bm{R},A)\cdot\bm{S}_{s}(\bm{R}+\bm{a}_{2},B)+J_{3}\bm{S}_{s}(\bm{R},A)\cdot\bm{S}_{s}(\bm{R}+\bm{a}_{3},B) ,\label{eq:bond_d} \\
&H_{\rm XY} 
=J_{-}\sum_{\bm{R}}
\bar{D}_{1}^{y}(S_{\rm u}^{x}(\bm{R},A)S_{\rm d}^{x}(\bm{R}+\bm{a}_{1},A)-S_{\rm u}^{y}(\bm{R},A)S_{\rm d}^{y}(\bm{R}+\bm{a}_{1},A)) \nonumber \\
&\hspace{18.5mm}+\bar{D}_{2}^{y}(S_{\rm u}^{x}(\bm{R},A)S_{\rm d}^{x}(\bm{R}+\bm{a}_{2},A)-S_{\rm u}^{y}(\bm{R},A)S_{\rm d}^{y}(\bm{R}+\bm{a}_{2},A)) \nonumber \\
&\hspace{18.5mm}+\bar{D}_{3}^{y}(S_{\rm u}^{x}(\bm{R},A)S_{\rm d}^{x}(\bm{R}+\bm{a}_{3},A)-S_{\rm u}^{y}(\bm{R},A)S_{\rm d}^{y}(\bm{R}+\bm{a}_{3},A)) \nonumber \\
&\hspace{18.5mm}+\bar{D}_{21}^{y}(S_{\rm u}^{x}(\bm{R},A)S_{\rm d}^{x}(\bm{R}+\bm{a}_{21},A)-S_{\rm u}^{y}(\bm{R},A)S_{\rm d}^{y}(\bm{R}+\bm{a}_{21},A)) \nonumber \\
&\hspace{18.5mm}+\bar{D}_{31}^{y}(S_{\rm u}^{x}(\bm{R},A)S_{\rm d}^{x}(\bm{R}+\bm{a}_{31},A)-S_{\rm u}^{y}(\bm{R},A)S_{\rm d}^{y}(\bm{R}+\bm{a}_{31},A)) \nonumber \\
&\hspace{18.5mm}+\bar{D}_{32}^{y}(S_{\rm u}^{x}(\bm{R},A)S_{\rm d}^{x}(\bm{R}+\bm{a}_{32},A)-S_{\rm u}^{y}(\bm{R},A)S_{\rm d}^{y}(\bm{R}+\bm{a}_{32},A)) \nonumber \\
&\hspace{18.5mm}-({\rm u}\leftrightarrow{\rm d}) \nonumber \\
&\hspace{18.5mm}-(A\leftrightarrow B) , \\
&H_{\rm \Gamma} 
=\Gamma\sum_{\bm{R}}
\bar{D}_{1}^{x}(S_{\rm u}^{x}(\bm{R},A)S_{\rm d}^{y}(\bm{R}+\bm{a}_{1},A)+S_{\rm u}^{y}(\bm{R},A)S_{\rm d}^{x}(\bm{R}+\bm{a}_{1},A)) \nonumber \\
&\hspace{14mm}+\bar{D}_{2}^{x}(S_{\rm u}^{x}(\bm{R},A)S_{\rm d}^{y}(\bm{R}+\bm{a}_{2},A)+S_{\rm u}^{y}(\bm{R},A)S_{\rm d}^{x}(\bm{R}+\bm{a}_{2},A)) \nonumber \\
&\hspace{14mm}+\bar{D}_{3}^{x}(S_{\rm u}^{x}(\bm{R},A)S_{\rm d}^{y}(\bm{R}+\bm{a}_{3},A)+S_{\rm u}^{y}(\bm{R},A)S_{\rm d}^{x}(\bm{R}+\bm{a}_{3},A)) \nonumber \\
&\hspace{14mm}+\bar{D}_{21}^{x}(S_{\rm u}^{x}(\bm{R},A)S_{\rm d}^{y}(\bm{R}+\bm{a}_{21},A)+S_{\rm u}^{y}(\bm{R},A)S_{\rm d}^{x}(\bm{R}+\bm{a}_{21},A)) \nonumber \\
&\hspace{14mm}+\bar{D}_{31}^{x}(S_{\rm u}^{x}(\bm{R},A)S_{\rm d}^{y}(\bm{R}+\bm{a}_{31},A)+S_{\rm u}^{y}(\bm{R},A)S_{\rm d}^{x}(\bm{R}+\bm{a}_{31},A)) \nonumber \\
&\hspace{14mm}+\bar{D}_{32}^{x}(S_{\rm u}^{x}(\bm{R},A)S_{\rm d}^{y}(\bm{R}+\bm{a}_{32},A)+S_{\rm u}^{y}(\bm{R},A)S_{\rm d}^{x}(\bm{R}+\bm{a}_{32},A)) \nonumber \\
&\hspace{14mm}-({\rm u}\leftrightarrow{\rm d}) \nonumber \\
&\hspace{14mm}-(A\leftrightarrow B) , \\
&H_{\kappa} =-\kappa \sum_{i,s={\rm u,d}}(S_{i,s}^{z})^{2},
\end{align}
where $\bm{S}_{i,{\rm u}}$ and $\bm{S}_{i,{\rm d}}$ are the operators of spins pointing upward and downward which are localized at the lattice site $i$, respectively.
Here, we write the spin operator $\bm{S}_{i,s}$ $(s={\rm u,d})$, in which the lattice site $i$ is the $X$ sublattice in the unit cell labeled by the lattice vector $\bm{R}$, as $\bm{S}_{s}(\bm{R},X)$ $(s={\rm u,d})$.
The vector $\bm{a}_{ij}$ is defined as  the difference between the lattice vectors $\bm{a}_{i}$ and $\bm{a}_{j}$, i.e.,  $\bm{a}_{ij}=\bm{a}_{i}-\bm{a}_{j}$.
We write the DM vectors $\bm{D}_{i}$ ($\bm{D}_{ij}$)  as $\bm{D}_{i}=D(\bm{d}^{1}_{i}(\bm{R})\times\bm{d}^{2}_{i}(\bm{R}))/|\bm{d}^{1}_{i}(\bm{R})\times\bm{d}^{2}_{i}(\bm{R})|$ $\left( \bm{D}_{ij}=D(\bm{d}^{1}_{ij}(\bm{R})\times\bm{d}^{2}_{ij}(\bm{R}))/|\bm{d}^{1}_{ij}(\bm{R})\times\bm{d}^{2}_{ij}(\bm{R})|\right)$.
Here, $\bm{d}^{1,2}_{i}(\bm{R})$ $(\bm{d}^{1,2}_{ij}(\bm{R}))$ are the two nearest neighbor bond vectors traversed between sites $(\bm{R},A)$ and $(\bm{R}+\bm{a}_{i},A)$ ($(\bm{R},A)$ and $(\bm{R}+\bm{a}_{ij},A)$)~\cite{Keffer62}.
Here $\bar{\bm{D}}_{i}$ and $\bar{\bm{D}}_{ij}$ are written as $\bar{\bm{D}}_{i}=\bm{D}_{i}/D$ and $\bar{\bm{D}}_{ij}=\bm{D}_{ij}/D$, respectively.

\noindent

\begin{table}[!h]
\caption{The topological invariants of the diamond lattice system. 
The parameters are chosen to be $J_{0}=1.4,J_{1}=J_{2}=J_{3}=J'=1.0,J_{-}=D=\Gamma=0.3,\kappa=1.5$. (see Ref.~\cite{Kondo19b} for definitions of the parameters).
The index $n=1,2$ denotes the upper band and the lower band in the particle space, respectively.
}
\begin{center}
{\tabcolsep=4mm
  \begin{tabular}{cccccccc}
    \hline
    $n$  & $\nu_{x,0}^{n}$ & $\nu_{x,\pi}^{n}$ &$\nu_{y,0}^{n}$  & $\nu_{y,\pi}^{n}$ & $\nu_{z,0}^{n}$ & $\nu_{z,\pi}^{n}$ & $(\nu_{0}^{n};\nu_{x}^{n},\nu_{y}^{n},\nu_{z}^{n})$  \\ \hline
    1  & 0 & 1& 0 & 1& 0 & 1 & (1;1,1,1)  \\ \hline
    2  & 0 & 1& 0 & 1& 0 & 1 & (1;1,1,1)  \\ \hline
  \end{tabular}
}
\end{center}
\label{table:index_di}
\end{table}
\noindent

Since this model has  the inversion symmetry, one can compute the topological invariants analytically by using a simplified formula (see Appendix B in Ref.~\cite{Kondo19b}) which can be thought of as the bosonic counterpart of the formula derived in Ref.~\cite{Fu07b}.
The Hamiltonian of the diamond lattice system satisfies the following inversion symmetry:
\begin{align}
R\Sigma_{z}H(\bm{k})-\Sigma_{z}H(-\bm{k})R=0,
\end{align}
where $R$ is an inversion operator defined as $R:=1_{2}\otimes 1_{2}\otimes \sigma_{x}$.
Following the discussion in Ref.~\cite{Fu07b}, topological invariants for 3D topological magnon systems with inversion symmetry can be written as
\begin{align}
&(-1)^{\nu_{0}}=\prod_{n_{1}=0,1;n_{2}=0,1;n_{3}=0,1}\delta_{m=(n_{1}n_{2}n_{3})}, \nonumber \\
&(-1)^{\nu_{i}}=\prod_{n_{i}=1;n_{j\neq i}=0,1}\delta_{m=(n_{1}n_{2}n_{3})}\label{eq:topo_inverse},
\end{align}
where $n_1, n_2, n_3 =0, 1$, and $i=x,y$, and $z$.
Since $\Sigma_{z}H(\bm{k})$ commutes with the inversion operator $R$ at TRIM: $\bm{\Gamma}_{m}=\pi(n_{1},n_{2},n_{3})$, an eigenvector $\bm{\Psi}_{n,1,+}(\bm{\Gamma}_{m})$ can be taken as an eigenvector of $R$. 
Here, we denote the eigenvalue of $R$ as $\xi_{n}(\bm{\Gamma}_{m})$. 
By the eigenvector, $\delta_{m=(n_{1}n_{2}n_{3})}$ in Eq.~(\ref{eq:topo_inverse}) is defined as the product of $\xi_{n}(\bm{\Gamma}_{m})$ over the bands below the virtual ``Fermi level" $\epsilon$
\begin{align}
\delta_{m=(n_{1}n_{2}n_{3})}=\prod_{n,(E_{n}({\bm k}) \leq \epsilon)}\xi_{n}(\bm{\Gamma}_{m}).\label{eq:delta}
\end{align}
From explicit expressions for the eigenvectors $ \bm{\Psi}_{n,1,+}({\bm \Gamma}_m) $, the strong index $ (-1)^{\nu_0} $ is obtained as
\begin{align}
(-1)^{\nu_{0}}
&={\rm sgn}[(J_{0}-J_{1}+J_{2}+J_{3})(J_{0}-J_{1}-J_{2}+J_{3})(J_{0}-J_{1}+J_{2}-J_{3})(J_{0}-J_{1}-J_{2}-J_{3})] \nonumber \\
&\times{\rm sgn}[(J_{0}+J_{1}+J_{2}+J_{3})(J_{0}+J_{1}-J_{2}+J_{3})(J_{0}+J_{1}+J_{2}-J_{3})(J_{0}+J_{1}-J_{2}-J_{3})].
\end{align}
Similarly, the other three indices are given as follows:
\begin{align}
&(-1)^{\nu_{x}}
={\rm sgn}[(J_{0}-J_{1}+J_{2}+J_{3})(J_{0}-J_{1}-J_{2}+J_{3})(J_{0}-J_{1}+J_{2}-J_{3})(J_{0}-J_{1}-J_{2}-J_{3})],
\\
&(-1)^{\nu_{y}}
={\rm sgn}[(J_{0}+J_{1}-J_{2}+J_{3})(J_{0}-J_{1}-J_{2}+J_{3})(J_{0}+J_{1}-J_{2}-J_{3})(J_{0}-J_{1}-J_{2}-J_{3})] ,
\\
&(-1)^{\nu_{z}}
={\rm sgn}[(J_{0}+J_{1}+J_{2}-J_{3})(J_{0}-J_{1}+J_{2}-J_{3})(J_{0}+J_{1}-J_{2}-J_{3})(J_{0}-J_{1}-J_{2}-J_{3})] .
\end{align}

\begin{figure}[!h]
\centering
  \includegraphics[width=14cm]{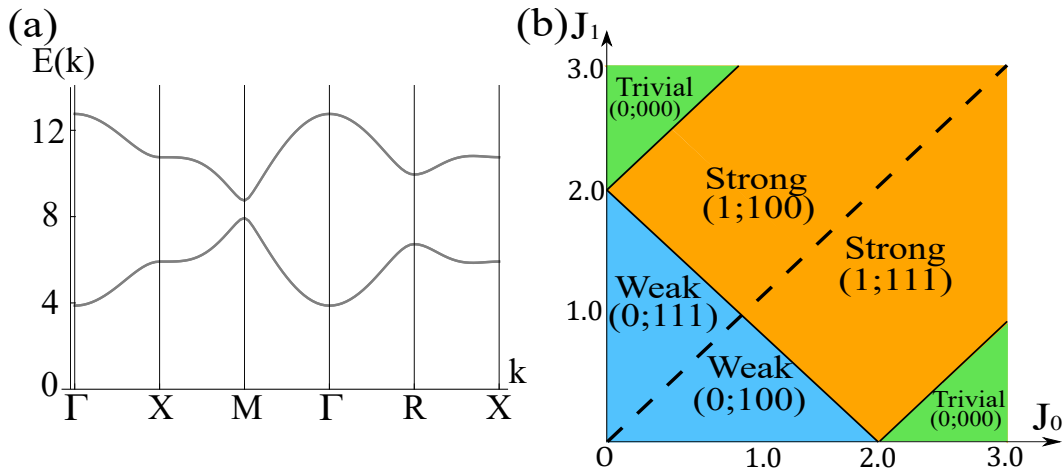}
\caption{(a) The bulk band structure of the diamond system expressed in terms of the Brillouin zone of the cubic lattice by taking $\bm{a}_{1}= (1,0,0), \bm{a}_{2} = (0,1,0)$, and $\bm{a}_{3} = (0,0,1)$ in Fig. 14.
Parameters are chosen to be $J_{0}=1.4$, $J_{1}=J_{2}=J_{3}=J'=1.0$, $J_{-}=D=\Gamma=0.3$, $\kappa=1.5$. The symmetry points are $\Gamma=(0,0,0)$, $X=(\pi,0,0)$, $M=(\pi,0,\pi)$, and $R=(\pi,\pi,\pi)$.
(b) The phase diagram of the diamond system as a function of $J_{0}$ and $J_{1}$. 
In each phase, the corresponding topological indices are for the lower band, indicated by $(\nu_{0}^{2};\nu_{x}^{2},\nu_{y}^{2},\nu_{z}^{2})$. 
The other parameters are chosen to be $J_{2}=J_{3}=J'=1.0$, $J_{-}=D=\Gamma=0.3$, $\kappa=1.5$.
The dashed line indicates the phase boundary between two phases with different weak indices, along which the energy gap vanishes.
The figures are taken from Ref.~\cite{Kondo19b}.
}\label{fig:bandPBC_di}
\end{figure}

Applying Eqs.~(\ref{eq:topo_inv}) and (\ref{eq:topo_inverse}) to the system, we obtain the set of the topological invariants summarized in Tab.~\ref{table:index_di}.
We have confirmed that the analytical results from Eq. (\ref{eq:delta}) of $(\nu_0^n;\nu_x^n,\nu_y^n,\nu_z^n)$ are exactly the same as those obtained by evaluating  Eq.~(\ref{eq:topo_inv}) numerically.
Table~\ref{table:index_di} suggests that the system is in the strong topological phase, i.e., an odd number of Dirac cones exist between the top and bottom bands in the particle (hole) space. 
The bulk band structure  under the periodic boundary condition with the same parameters as those of Table~\ref{table:index_di} is shown in Fig.~\ref{fig:bandPBC_di}(a). Since the system has both the pseudo-time-reversal symmetry and inversion symmetry, each band is doubly degenerate over the whole Brillouin zone.

Using the simplified formula Eq.~(\ref{eq:topo_inverse}), we analytically construct a phase diagram of the diamond lattice system drawn in Fig.~\ref{fig:bandPBC_di}(b). As shown there, three topologically distinct phases: the strong, weak, and trivial phases are all realized in this system. We note that from the numerical calculations, the band gap seems to close only at TRIM, thus we were able to draw the phase diagram analytically by Eq.~(\ref{eq:topo_inverse}).

\begin{figure}[!h]
\centering
  \includegraphics[width=14cm]{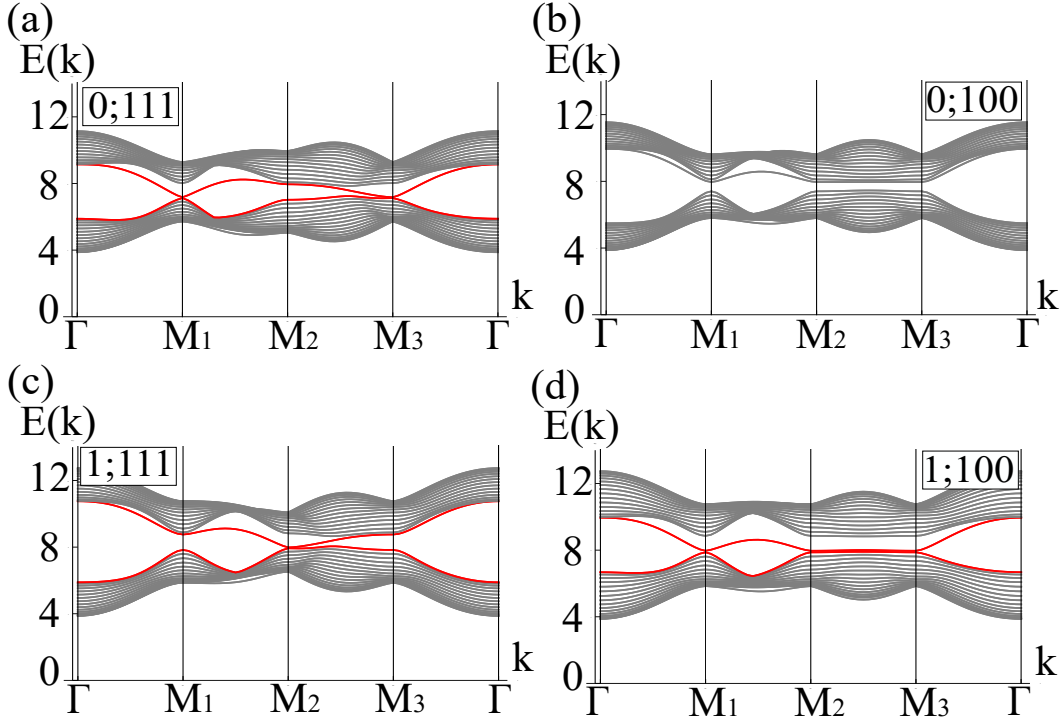}
\caption{Band structures for a slab with a (100) face for (a), (b) the weak and (c), (d) the strong topological phases in Fig.~\ref{fig:bandPBC_di}(b).
The red lines indicate the surface states protected by pseudo-time-reversal symmetry.
The symmetry points are $\Gamma=(0,0),M_{1}=(\pi,0),M_{2}=(\pi,\pi)$, and $M_{3}=(0,\pi)$.
Coupling constants $(J_{0},J_{1})$ are chosen to be (a) $(J_{0},J_{1})=(0.6, 1.0)$,  (b) $(J_{0},J_{1})=(1.0, 0.8)$,  (c) $(J_{0},J_{1})=(1.4, 1.0)$, and  (d) $(J_{0},J_{1})=(1.0, 1.4)$, respectively. The other parameters are the same as those in Fig.~\ref{fig:bandPBC_di}(b).
The figures are taken from Ref.~\cite{Kondo19b}.
}\label{fig:bandOBC_di}
\end{figure}

Figure~\ref{fig:bandOBC_di} shows band structures for a slab with a (100) face for the four phases in Fig.~\ref{fig:bandPBC_di}(b). 
As expected from the general discussion in Sec.~\ref{e_3DTI}, in weak topological phases with $(\nu_{0};\nu_{x},\nu_{y},\nu_{z})=(0;111)$ of Fig.~\ref{fig:bandOBC_di}(a) and $(0;100)$ of Fig.~\ref{fig:bandOBC_di}(b), there are even number (2 and 0, respectively) of Dirac cones. On the other hand, in strong topological phases with $(1; 111)$ of Fig.~\ref{fig:bandOBC_di}(c) and $(1; 100)$ of Fig.~\ref{fig:bandOBC_di}(d), there are odd number (1 and 3, respectively) of Dirac cones.
We note in passing that other examples of 3D topological magnon systems are provided in Ref.~\cite{Kondo19b}.
The analysis of these systems is more involved than that of the diamond lattice system since they lack inversion symmetry and the set of topological invariants has to be computed numerically by using Eq.~(\ref{eq:topo_inv}).

\noindent

\noindent

\subsubsection{The thermal Hall effect on the surface of 3D topological magnon systems\label{surf_Hall}}

\begin{figure}[!h]
\centering
  \includegraphics[width=14cm]{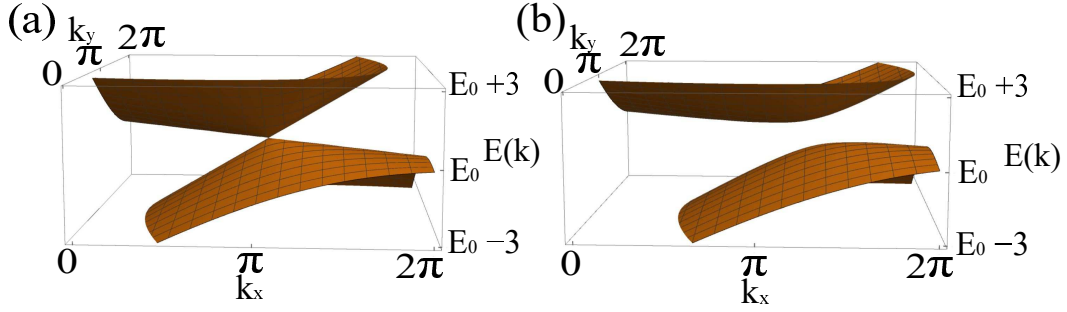}
\caption{Band structure of the single surface state of the diamond lattice system (a) without and (b) with the surface magnetic field $BS=1.0$, where $E_{0}=8.34$ is the band touching energy.
Other parameters are chosen to be  $J_{0}S=1.4,J_{1}S=J_{2}S=J_{3}S=J'S=1.0,J_{-}S=DS=\Gamma S=0.3, \kappa S=1.5$. 
The figures are taken from Ref.~\cite{Kondo19b}.
}\label{fig:diamond_surface}
\end{figure}

\begin{figure}[!h]
\centering
  \includegraphics[width=14cm]{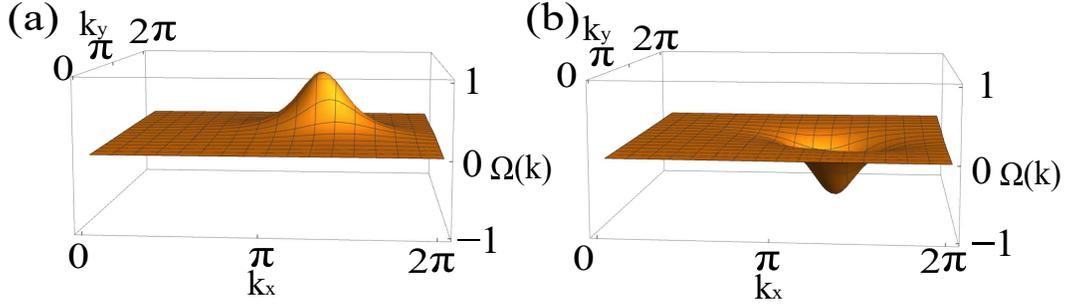}
\caption{Berry curvatures of  (a) the top and (b) the bottom bands of the surface state in Fig.~\ref{fig:diamond_surface}(b).
Parameters are taken as $J_{0}S=1.4,J_{1}S=J_{2}S=J_{3}S=J'S=1.0,J_{-}S=DS=\Gamma S=0.3, \kappa S=1.5, BS=1.0$.
}\label{fig:diamond_surface_Berry}
\end{figure}

In this part, we discuss the physical implication of the surface state in the strong topological phase --- the thermal Hall effect on the surface.
Previous studies~\cite{Sinitsyn07,Lu10} showed that the Dirac dispersion in the surface states in 3D strong topological insulators in class AII can be gapped out by applying a magnetic field to the surface due to the breaking of time-reversal symmetry.
In general, such states are shown to have nonzero Berry curvature, giving rise to the surface quantum Hall effect.
The analogous effect is expected to occur in 3D topological magnon systems as discussed in Ref.~\cite{Kondo19b}.
In this case, the effective Hamiltonian for the surface states can be written as follows:
\begin{align}
H_{\rm eff}(k_{x},k_{y})=\left(
\begin{array}{cc}
\bra{\bm{\psi}}H_{xy}(k_{x},k_{y})\ket{\bm{\psi}}+E_{0} & \bra{\bm{\psi}}H_{xy}(k_{x},k_{y})\ket{\Theta'\bm{\psi}} \\
\bra{\Theta'\bm{\psi}}H_{xy}(k_{x},k_{y})\ket{\bm{\psi}} &\bra{\Theta'\bm{\psi}}H_{xy}(k_{x},k_{y})\ket{\Theta'\bm{\psi}}+E_{0} 
\end{array}
\right),\label{eq:Heff}
\end{align}
where the matrix element $\bra{\bm{\psi}}H_{xy}(k_{x},k_{y})\ket{\bm{\psi}}$ is defined as
\begin{align}
\bra{\bm{\psi}}H_{xy}(k_{x},k_{y})\ket{\bm{\psi}}=\int_{-\infty}^{0}dz\bm{\psi}^{\dagger}(z)H_{xy}(k_{x},k_{y})\bm{\psi}(z).
\end{align}
Here $\bm{\psi}(z)$ is a wave function of the surface Dirac states.
The matrix $H_{xy}(k_{x},k_{y})$ is the first order term in $k_{x}$ and $k_{y}$. Here $E_{0}$ is the eigenenergy of the surface Dirac states.
The other three matrix elements are defined similarly.
By applying the magnetic field $\bm{B}=B\bm{e}_{z}$, on the surface by making, for instance, a heterostructure of a ferromagnet and the 3D topological magnon system, the additional term $-BS\sigma_{z}$ appears in the Hamiltonian Eq.~(\ref{eq:Heff}).
The band structure of the surface is obtained by diagonalizing the Hamiltonian $H_{\rm eff}(k_{x},k_{y})$ by a unitary matrix.
Note that the effective Hamiltonian is Hermitian since only the states in the particle space are involved.
The Berry curvature of the system is defined as $\Omega_{n}^{z}(k_{x},k_{y})=2{\rm Im}\left[(\partial_{k_{x}}\bm{\psi}_{n}^{\dagger}(k_{x},k_{y}))(\partial_{k_{y}}\bm{\psi}_{n}(k_{x},k_{y}))\right]$, where $n=1$ and $n=2$ indicate the upper and lower bands of the surface Dirac states, respectively.
Figure~\ref{fig:diamond_surface} shows the band structure of the surface states without and with the surface magnetic field. 
As it is clear, the Dirac dispersion can be gapped out by applying the surface magnetic field.
Figure~\ref{fig:diamond_surface_Berry} shows the corresponding Berry curvatures of the top and bottom bands under the magnetic field.
Thanks to the nonvanishing Berry curvature, the thermal Hall coefficient Eq.~(\ref{eq:thermal_Hall_coefficient}) is expected to be nonzero.

\noindent
\noindent

\section{Summary\label{summary}}

In this paper, we reviewed the recent development in the study of topological phases of magnon BdG systems.
Fermion-like pseudo-time-reversal symmetry we introduced plays an important role in constructing magnonic counterparts of class AII topological insulators in 2D and 3D.
As a major difference from fermionic systems, bosonic BdG systems have a unique mathematical property –- non-Hermiticity.
Therefore, the bosonic BdG systems are categorized by the topological classification of non-Hermitian systems.

As a first step to construct symmetry-protected topological magnon phases, we have introduced the nontrivial pseudo-time-reversal symmetry which ensures the presence of Kramers pairs of magnons (bosons).
Then, we identified the ${\mathbb Z}_{2}$ topological invariant which characterizes the 2D magnon spin Hall systems.
To demonstrate the validity of the invariant, we constructed and studied two models of magnon spin Hall systems, the bilayer kagome and honeycomb systems. 
In both cases, we confirmed numerically that the ${\mathbb Z}_{2}$ index characterizes the presence of edge states and remains robust against small changes in the parameters.
The latter, bilayer honeycomb system, is the first model of magnon spin Hall systems without the spin conservation, and can be thought of as a magnonic analog of the Kane-Mele model~\cite{Kane05a,Kane05b}. 
In addition, generalizing the former system, we found that bilayer CrI$_3$ is a candidate material for realizing the magnon spin Hall system.

We extended the idea of the above magnon spin Hall systems to 3D systems, giving a specific model on the diamond lattice.
The model in 3D also has pseudo-time-reversal symmetry, where we can define the set of topological invariants.
Thanks to the additional symmetry, i.e., inversion symmetry of the model, we simplified the formula of the topological invariants. 
The simplified formula allows us to compute the topological invariants analytically and draw the phase diagram including the strong topological, weak topological, and trivial phases, in which the number of surface Dirac states is odd, even, and zero, respectively.
In addition, as a physical consequence of the single surface Dirac dispersion in the strong topological phase, we predicted that the thermal Hall effect of surface magnons occurs in the presence of a magnetic field due to the proximity to a normal ferromagnet.

We here emphasize that the ${\mathbb Z}_{2}$ topological invariants defined in terms of the bosonic Berry connection and curvature are applicable to other bosonic systems such as phonons and photons, as long as they respect pseudo-time-reversal symmetry. Relatedly, it would also be interesting to study bosonic excitations in spin liquids or paramagnets by combining our approach with the Schwinger-boson mean-field theory~\cite{Lee14}.
To construct other symmetry-protected topological magnon phases, such as magnonic analog of topological crystalline insulators~\cite{Fu11} would be one of the future directions.
Last but not least, since various methods for measuring the magnon current, accumulation, and so on have been developed~\cite{Onose10,Hirschberger15,Fukuhara13}, we expect the magnon surface states and the related phenomena to be observed in real materials in the near future.
We hope that our work will stimulate further studies on magnon topological phases.

\section*{Acknowledgment}


H. Katsura thanks Patrick A. Lee, Naoto Nagaosa, and Yoshinori Onose, for fruitful discussions and collaborations. 
He also thanks Dahlia Klein for an inspiring discussion.
H. Kondo and Y. Akagi thank Kohei Kawabata for insightful comments.
This work was supported by JSPS KAKENHI Grants No. JP17K14352, JP18K03445, No. JP20K14411, No. JP20J12861, and JSPS Grant-in-Aid for Scientific Research on Innovative Areas “Topological Materials Science” (KAKENHI Grant No. JP18H04220), “Discrete Geometric Analysis for Materials Design” (KAKENHI Grant No. JP20H04630), and “Quantum Liquid Crystals” (KAKENHI Grant No. JP20H05154) from JSPS of Japan. 
H. Kondo was supported by the  JSPS through Program for Leading Graduate Schools (ALPS).
H. Katsura was supported by the Inamori Foundation.


%

\appendix

\section{Proof of the statements in Sec.~\ref{boson_BdG}}

In this Appendix, we prove the three statements (i)-(iii) in Sec.~\ref{boson_BdG}.

\noindent
{\it Proof of (i).}

\noindent

If the matrix $H(\bm{k})$ is positive definite, $H(\bm{k})$ can be written as follows:
\begin{align}
H(\bm{k})=U^{\dagger}(\bm{k}){\rm diag}(\lambda_{1}(\bm{k}),\cdots,\lambda_{2\mathscr{N}}(\bm{k}))U(\bm{k}),
\end{align}
where $U(\bm{k})$ is a unitary matrix and $\lambda_{n}(\bm{k})$ $(n=1,\cdots,2\mathscr{N})$ are the eigenvalues of $H(\bm{k})$. By using a regular matrix defined as $Q(\bm{k})={\rm diag}(\sqrt{\lambda_{1}(\bm{k})},\cdots,\sqrt{\lambda_{2\mathscr{N}}(\bm{k})})U(\bm{k})$, $H(\bm{k})$ is written as $H(\bm{k})=Q^{\dagger}(\bm{k})Q(\bm{k})$.
Since the matrix $\Sigma_{z}H(\bm{k})=\Sigma_{z}Q^{\dagger}(\bm{k})Q(\bm{k})$ is similar to the matrix $\Omega ({\bm k})=Q(\bm{k})\Sigma_{z}Q^{\dagger}(\bm{k})$, $\Sigma_{z}H(\bm{k})$ has the same eigenvalues as $\Omega ({\bm k})$.
On the other hand, the eigenvalues of the matrix $\Omega ({\bm k})$ is real and nonzero since it is Hermitian and satisfies
\begin{align}
{\rm Det}\left(\Omega(\bm{k})\right)={\rm Det}\left(\Sigma_{z}\right){\rm Det}\left(H(\bm{k})\right)
=(-1)^{\mathscr{N}}{\rm Det}\left(Q(\bm{k})Q^{\dagger}(\bm{k})\right) \neq 0.
\end{align}
The above leads to the conclusion that the eigenvalues of the matrix $\Sigma_{z}H(\bm{k})$ are real and nonzero.

\qed

\noindent{\it Proof of (ii).}
\noindent
Multiplying the complex conjugate of the eigen-equation $\Sigma_{z}H(\bm{k})\bm{\psi}(\bm{k})=E(\bm{k})\bm{\psi}(\bm{k})$ from the left by $\Sigma_{x}$ and reversing the direction of the wave vector, we obtain
\begin{align}
-\Sigma_{z}\Sigma_{x}H^{*}(-\bm{k})\Sigma_{x}\Sigma_{x}\bm{\psi}^{*}(-\bm{k})=E(-\bm{k})\Sigma_{x}\bm{\psi}^{*}(-\bm{k}),
\end{align}
where we used the anti-commutation relation $\{ \Sigma_{z},\Sigma_{x}\}=0$ and $\Sigma_{x}^{2}=1_{2\mathscr{N}}$.
By using the relation $\Sigma_{x}H(\bm{k})\Sigma_{x}=H^{*}(-\bm{k})$, the above equation can be rewritten as 
\begin{align}
\Sigma_{z}H(\bm{k})\Sigma_{x}\bm{\psi}^{*}(-\bm{k})=-E(-\bm{k})\Sigma_{x}\bm{\psi}^{*}(-\bm{k}).
\end{align}

\qed

\noindent
Here, we can arrange the eigenvalues and the eigenvectors of $\Sigma_{z}H(\bm{k})$ as
\begin{align}
&\left(E_{1}(\bm{k}),\cdots,E_{\mathscr{N}}(\bm{k}),-E_{1}(-\bm{k}),\cdots,-E_{\mathscr{N}}(-\bm{k})\right),\\
&\left(\bm{\psi}_{1}(\bm{k}),\cdots,\bm{\psi}_{\mathscr{N}}(\bm{k}),\Sigma_{x}\bm{\psi}^{*}_{1}(-\bm{k}),\cdots,\Sigma_{x}\bm{\psi}^{*}_{\mathscr{N}}(-\bm{k})\right).\label{eq:eigenvectors}
\end{align}
For later convenience, the eigenvectors are denoted by
\begin{align}
&\bm{\psi}_{n+}(\bm{k})=\bm{\psi}_{n}(\bm{k}), \\
&\bm{\psi}_{n-}(\bm{k})=\Sigma_{x}\bm{\psi}^{*}_{n}(-\bm{k}).
\end{align}
We note in passing that Eq.~(\ref{eq:eigenvectors}) turns out to be the para-unitary matrix $T(\bm{k})$ in Eq. (28).

\noindent{\it Proof of (iii).}
\noindent

Let us begin with the eigenequation of the matrix $\Omega(\bm{k})=Q(\bm{k})\Sigma_{z}Q^{\dagger}(\bm{k})$:
\begin{align}
\Omega(\bm{k})\bm{\phi}_{n\sigma}(\bm{k})=\sigma E_{n}(\bm{k})\bm{\phi}_{n\sigma}(\bm{k}) \hspace{10mm}(E_{n}(\bm{k})>0).
\end{align}
Since $\Omega(\bm{k})$ is Hermitian, the eigenvectors $\bm{\phi}_{n\sigma}(\bm{k})$ can be chosen to be orthonormal, i.e., 
\begin{align}
\langle \bm{\phi}_{m\rho}(\bm{k}),\bm{\phi}_{n\sigma}(\bm{k})\rangle=\bm{\phi}_{m\rho}^{\dagger}(\bm{k})\bm{\phi}_{n\sigma}(\bm{k})=\delta_{nm}\delta_{\sigma\rho}.
\end{align}
We now define the vectors as $\bm{\psi}_{n\sigma}(\bm{k}):=\sqrt{E_{n}(\bm{k})}Q^{-1}(\bm{k})\bm{\phi}_{n\sigma}(\bm{k})$ which satisfy
\begin{align}
\Sigma_{z}H(\bm{k})\bm{\psi}_{n\sigma}(\bm{k})=\sigma E_{n}(\bm{k})\bm{\psi}_{n\sigma}(\bm{k}).
\end{align}
Thus,  $\bm{\psi}_{n\sigma}(\bm{k})$ is an eigenvector of the matrix $\Sigma_{z}H(\bm{k})$ with eigenvalue $\sigma E_{n}(\bm{k})$.
The vector $\bm{\psi}_{n\sigma}(\bm{k})$ satisfies the following para-unitarity relation:
\begin{align}
&\langle\langle \bm{\psi}_{m\rho}(\bm{k}),\bm{\psi}_{n\sigma}(\bm{k})\rangle\rangle \nonumber \\
&=\bm{\psi}_{m\rho}^{\dagger}(\bm{k})\Sigma_{z}\bm{\psi}_{n\sigma}(\bm{k})\nonumber \\
&=\sqrt{E_{n}(\bm{k})E_{m}(\bm{k})}\bm{\phi}_{m\rho}^{\dagger}(\bm{k})[Q^{-1}(\bm{k})]^{-1}\Sigma_{z} Q^{-1}(\bm{k})\bm{\phi}_{n\sigma}(\bm{k})\nonumber \\
&=\sigma\sqrt{\frac{E_{n}(\bm{k})}{E_{m}(\bm{k})}}\bm{\phi}_{m\rho}^{\dagger}(\bm{k})\bm{\phi}_{n\sigma}(\bm{k})\nonumber \\
&=\sigma\delta_{nm}\delta_{\sigma\rho}.
\end{align}
In the third equality, we used the following equation: 
\begin{align}
&[Q^{-1}(\bm{k})]^{-1}\Sigma_{z} Q^{-1}(\bm{k})\bm{\phi}_{n\sigma}(\bm{k})\nonumber \\
&=\Omega^{-1}(\bm{k})\bm{\phi}_{n\sigma}(\bm{k})\nonumber \\
&=(\sigma E_{n}(\bm{k}))^{-1}\bm{\phi}_{n\sigma}(\bm{k}).
\end{align}

\qed

\section{Berry connection/curvature and spin Chern number of bilayer ``ferromagnet"\label{bi_ferro}}

In Sec.~\ref{model_2D}, we provided the bilayer kagome ``ferromagnet" as an example of a magnon spin Hall system which is characterized by the ${\mathbb Z}_{2}$ topological invariant Eq. (\ref{eq:D}) or spin Chern number. 
In this Appendix, we extend the model to include more general bilayer "ferromagnetic" systems without a net moment. 
We here assume that every single layer has a nonzero Chern number and combine the two single layers so that the total bilayer system restores (pseudo-)time-reversal symmetry.
At the end of the Appendix, we will show that the Berry connections and curvatures of the bilayer system perfectly coincide with those of the two independent single layer systems without the interlayer coupling $J'$, which means that  the bilayer system is characterized by nonzero spin Chern number.

The BdG Hamiltonian of the ``ferromagnetic'' bilayer system 
takes the same form as Eq.~(\ref{eq:H}), i.e., 
\begin{align}
H(\bm{k})=
\left(
\begin{array}{cccc}
H^{(\rm{single})}(\bm{k}) &0 &0 &J'S1_{N} \\
0 &H^{(\rm{single})*}(-\bm{k})  &J'S1_{N}&0 \\
0 &J'S1_{N} &H^{(\rm{single})*}(-\bm{k}) &0 \\
J'S1_{N} &0 &0 &H^{(\rm{single})}(\bm{k}) \\
\end{array} 
\right)+J'S1_{4N},
\end{align}
where $H^{(\rm{single})}(\bm{k})$ is the Hamiltonian of the ferromagnetic single layer system.
To diagonalize the Hamiltonian with the para-unitary matrix $T(\bm{k})$ which satisfies the condition $T^{\dagger}(\bm{k})\Sigma_{z}T(\bm{k})=\Sigma_{z}$, we need to solve the eigenvalue problem:
\begin{align}
\Sigma_{z}H(\bm{k})\bm{\Psi}_{n,l,\sigma}(\bm{k})=E_{n,l,\sigma}(\bm{k})\bm{\Psi}_{n,l,\sigma}(\bm{k}). 
\label{eq:BdG_GF}
\end{align}
Thanks to the particular block structure of $H({\bm k})$, the eigenvectors $\bm{\Psi}_{n,l,\sigma} ({\bm k})$ can be constructed from the eigenvectors of the single-layer Hamiltonian with particle-number conservation. Denoting by $\bm{\psi}_n ({\bm k})$ the eigenvector of $H^{({\rm single})} ({\bm k})$ with eigenvalue $\lambda_n ({\bm k})$, the corresponding eigenvalues and eigenvectors of $\Sigma_z H({\bm k})$ read
\begin{align}
&E_{n,1,\sigma}(\bm{k})=\sigma\sqrt{\left({\lambda}_{n}(\sigma\bm{k})+J'S\right)^2-(J'S)^2},  \\
&E_{n,2,\sigma}(\bm{k})=\sigma\sqrt{\left({\lambda}_{n}(-\sigma\bm{k})+J'S\right)^2-(J'S)^2}, \\
&\bm{\Psi}_{n,1,+}(\bm{k})=
\left(
\begin{array}{cccc}
\cosh{(\theta_{n}(\bm{k}))}\bm{\psi}_{n}(\bm{k}) \\
0  \\
0  \\
\sinh{(\theta_{n}(\bm{k}))}\bm{\psi}_{n}(\bm{k}) \\
\end{array} 
\right),  \label{eq:1+}\\
&\bm{\Psi}_{n,2,+}(\bm{k})=
\left(
\begin{array}{cccc}
0 \\
\cosh{(\theta_{n}(-\bm{k}))}\bm{\psi}_{n}^{*}(-\bm{k})  \\
\sinh{(\theta_{n}(-\bm{k}))}\bm{\psi}_{n}^{*}(-\bm{k})  \\
0 
\end{array} 
\right),  \label{eq:2+}\\
&\bm{\Psi}_{n,1,-}(\bm{k})=
\left(
\begin{array}{cccc}
0 \\
\sinh{(\theta_{n}(-\bm{k}))}\bm{\psi}_{n}^{*}(-\bm{k})  \\
\cosh{(\theta_{n}(-\bm{k}))}\bm{\psi}_{n}^{*}(-\bm{k})  \\
0 
\end{array} 
\right), \label{eq:1-}\\
&\bm{\Psi}_{n,2,-}(\bm{k})=
\left(
\begin{array}{cccc}
\sinh{(\theta_{n}(\bm{k}))}\bm{\psi}_{n}(\bm{k}) \\
0  \\
0  \\
\cosh{(\theta_{n}(\bm{k}))}\bm{\psi}_{n}(\bm{k}) 
\end{array} 
\right), \label{eq:2-}
\end{align}
where $\theta_{n}(\bm{k})$ is defined by
\begin{align}
\tanh{\left(\theta_{n}(\bm{k})\right)}=\frac{-\left(\lambda_{n}(\bm{k})+J'S\right)+\sqrt{\left(\lambda_{n}(\bm{k})+J'S\right)^2-(J'S)^2}}{J'S}.
\end{align}
Then the para-unitary matrix $T(\bm{k})$ and the diagonalized Hamiltonian are 
given by
\begin{align}
&T(\bm{k})=\left(
\Phi_{1,+}(\bm{k}),\Phi_{2,+}(\bm{k}),\Phi_{1,-}(\bm{k}),\Phi_{2,-}(\bm{k})\right), \\
&T^{\dagger}(\bm{k})H(\bm{k})T(\bm{k}) =\left(
\begin{array}{cccc}
E_{1,+}(\bm{k}) &0 &0 &0 \\
0 &E_{2,+}(\bm{k}) &0 &0  \\
0 &0 &-E_{1,-}(\bm{k}) &0  \\
0 &0 &0 &-E_{2,-}(\bm{k})
\end{array} 
\right).
\end{align}
Here, a $4N\times N$ matrix $\Phi_{l,\sigma}(\bm{k})$ and an $N\times N$ diagonal matrix $E_{l,\sigma}(\bm{k})$ is defined as
\begin{align}
&\Phi_{l,\sigma}(\bm{k})=\left[\bm{\Psi}_{1,l,\sigma}(\bm{k}),\cdots\!,\bm{\Psi}_{N,l,\sigma}(\bm{k})\right] ,\\
&E_{l,\sigma}(\bm{k})={\rm diag}\left[ E_{1,l,\sigma}(\bm{k}),\cdots,E_{N,l,\sigma}(\bm{k})\right].
\end{align}
By substituting Eqs.~(\ref{eq:1+})-(\ref{eq:2-}) into Eq.~(\ref{eq:BdGBerryconnection}), 
we find the following relations for the Berry connection:
\begin{align}
&\bm{A}_{n,1,+}(\bm{k}) =\bm{A}_{n,2,-}(\bm{k})\nonumber \\
&={\rm i}\,\left\langle\cosh{(\theta_{n}(\bm{k}))}\bm{\psi}_{n}(\bm{k}),
\nabla_{\bm{k}}\cosh{(\theta_{n}(\bm{k}))}\bm{\psi}_{n}(\bm{k})\right\rangle
- {\rm i}\,\left\langle\sinh{(\theta_{n}(\bm{k}))}\bm{\psi}_{n}(\bm{k}),
\nabla_{\bm{k}}\sinh{(\theta_{n}(\bm{k}))}\bm{\psi}_{n}(\bm{k})\right\rangle \nonumber \\
&={\rm i}\,\left\langle\bm{\psi}_{n}(\bm{k}),\nabla_{\bm{k}}\bm{\psi}_{n}(\bm{k})\right\rangle\nonumber \\
&=\bm{A}_{n}^{(\rm{single})}(\bm{k}), \label{eq:Berryconnection_bilayer_singlelayer}
\\
&\bm{A}_{n,2,+}(\bm{k}) =\bm{A}_{n,1,-}(\bm{k})\nonumber \\
&={\rm i}\,\left\langle\cosh{(\theta_{n}(-\bm{k}))}\bm{\psi}_{n}(-\bm{k}),
\nabla_{\bm{k}}\cosh{(\theta_{n}(-\bm{k}))}\bm{\psi}_{n}(-\bm{k})\right\rangle^{*}\nonumber \\
&\hspace{6mm}-{\rm i}\,\left\langle\sinh{(\theta_{n}(-\bm{k}))}\bm{\psi}_{n}(-\bm{k}),
\nabla_{\bm{k}}\sinh{(\theta_{n}(-\bm{k}))}\bm{\psi}_{n}(-\bm{k})\right\rangle^{*}\nonumber \\
&={\rm i}\,\left\langle\bm{\psi}_{n}(-\bm{k}),\nabla_{\bm{k}}\bm{\psi}_{n}(-\bm{k})\right\rangle^{*} \nonumber \\
&={\rm i}\,\left\langle\bm{\psi}_{n}(-\bm{k}),\nabla_{-\bm{k}}\bm{\psi}_{n}(-\bm{k})\right\rangle\nonumber \\
&=\bm{A}_{n}^{(\rm{single})}(-\bm{k}),
\end{align}
where $\bm{A}_{n}^{(\rm{single})}(\bm{k})={\rm i}\,\left\langle\bm{\psi}_{n}(\bm{k}),\nabla_{\bm{k}}\bm{\psi}_{n}(\bm{k})\right\rangle$ is the Berry connection of the single layer system.
Using the Berry curvature of the single layer system 
 $\Omega_{n}^{({\rm single})}(\bm{k})=\bigl( \nabla_{\bm{k}} \times \bm{A}_{n}^{(\rm{single})}(\bm{k}) \bigr)_z$, the Berry curvature (\ref{eq:BdGBerrycurvature}) can be written as
\begin{align}
&\Omega_{n,1,+}^{z}(\bm{k})=\Omega_{n,2,-}^{z}(\bm{k})=\Omega_{n}^{(\rm{single})}(\bm{k}), \nonumber \\
&\Omega_{n,2,+}^{z}(\bm{k})=\Omega_{n,1,-}^{z}(\bm{k})=-\Omega_{n}^{(\rm{single})}(-\bm{k}). \label{eq:Berry_bilayer_singlelayer}
\end{align}
As seen in Eqs. (\ref{eq:Berryconnection_bilayer_singlelayer}) - (\ref{eq:Berry_bilayer_singlelayer}), the Berry connection and curvature of a “ferromagnetic” bilayer system can be written in terms of those of the two independent single layer systems without the interlayer coupling $J'$.
Since we assumed that each single layer system we considered here is characterized by a nonzero Chern number given by ${\rm C}_{n,l}=\frac{1}{2\pi}\int_{\rm BZ} d^2k \Omega_{n,l,+}(\bm{k})$, the total bilayer system exhibits magnon spin Hall effect due to the nonzero spin Chern number $D_n=({\rm C}_{n,1}-{\rm C}_{n,2})/2$.
We here emphasize that this argument is valid in general ``ferromagnetic" bilayer systems where spins on the same layer point in the same direction while spins on different layers point in opposite directions.
Therefore, we can simply construct the magnon spin Hall systems by combining two single layers, each of which exhibits the thermal Hall effect.

\end{document}